\documentstyle[myart,amsfonts,12pt]{article}

\normalbaselines
\font\tenbm=cmmib10
\font\sevenbm=cmmib7
\font\fivebm=cmmib5
\newfam\bmfam
\textfont\bmfam=\tenbm \scriptfont\bmfam=\sevenbm
\scriptscriptfont\bmfam=\fivebm
{\count0=\number\bmfam \multiply\count0 by "100
\def\defbgreek#1#2#3{{\count1=\count0 \advance\count1 by "#2#3
  \global\mathchardef#1=\count1 }}
\defbgreek\balpha  0B \defbgreek\brho       1A
\defbgreek\bbeta   0C \defbgreek\bsigma     1B
\defbgreek\bgamma  0D \defbgreek\btau       1C
\defbgreek\bdelta  0E \defbgreek\bupsilon   1D
\defbgreek\bepsilon0F \defbgreek\bphi       1E
\defbgreek\bzeta   10 \defbgreek\bchi       1F
\defbgreek\bmeta   11 \defbgreek\bpsi       20
\defbgreek\btheta  12 \defbgreek\bomega     21
\defbgreek\biota   13 \defbgreek\bvarepsilon22
\defbgreek\bkappa  14 \defbgreek\bvartheta  23
\defbgreek\blambda 15 \defbgreek\bvarpi     24
\defbgreek\bmu     16 \defbgreek\bvarrho    25
\defbgreek\bnu     17 \defbgreek\bvarsigma  26
        \defbgreek\bxi     18 \defbgreek\bvarphi    27
\defbgreek\bpi     19}
\input tcilatex.tex

\begin{document}

\author{Yuri A. Rylov}
\title{Associative delusions and problem of their overcoming}
\date{Institute for Problems in Mechanics, Russian Academy of Sciences \\
101-1 ,Vernadskii Ave., Moscow, 117526, Russia \\
email: rylov@ipmnet.ru\\
Web site: {$http://rsfq1.physics.sunysb.edu/\symbol{126}rylov/yrylov.htm$}\\
or mirror Web site: {$http://194.190.131.172/\symbol{126}rylov/yrylov.htm$}}
\maketitle

\begin{abstract}
Influence of associative delusions (AD) onto development of physics and
mathematics is investigated. The associative delusion (AD) means a mistake,
appearing from incorrect associations, when a property of one object is
attributed to another one. Examples of most ancient delusions are:
(1)~connection of the gravitation field direction with a preferred direction
in space (instead of the direction to the Earth centre), that had lead to
the antipode paradox, (2)~statement that the Earth (not the Sun) is the
centre of the planetary system, that had lead to the Ptolemaic doctrine. Now
these ADs have been overcame. In the paper one considers four modern and not
yet got over ADs, whose corollaries are false space-time geometry in the
micro world and most of problems and difficulties of the quantum field
theory (QFT). One shows that ADs have a series of interesting properties:
(1)~ADs appear to be long-living delusions, because they are compensated
partly by means of introduction of compensating (Ptolemaic) conceptions,
(2)~ADs influence on scientific investigations, generating a special
pragmatic style (P-style) of investigations resembling experimental trial
and error method, (3)~acting on investigations directly and via P-style, ADs
direct the science development into a blind alley. One considers concrete
properties of modern ADs and the methods of their overcoming. From viewpoint
of application the paper is an analysis of mistakes, made in the quantum
theory development. One analyses reasons of these mistakes and suggests
methods of their correction.
\end{abstract}

\newpage

\section{Introduction}

The present paper is devoted to a study of associative delusions, their role
in the natural science development and to problems of their overcoming. The
associative delusion means such a situation, when associative properties of
human thinking actuate incorrectly, and the natural phenomenon is attributed
by properties alien to it. Usually one physical phenomenon is attributed by
properties of other physical phenomenon, or properties of the physical
phenomenon description are attributed to the physical phenomenon in itself.
Let us illustrate this in a simple example, which is perceived now as a
grotesque.

It is known that ancient Egyptians believed that all rivers flow towards the
North. This delusion seems now to be nonsense. But many years ago it had
weighty foundation. The ancient Egyptians lived on a vast flat plane and
knew only one river the Nile, which flowed exactly towards the North and had
no tributaries on the Egyptian territory. The North direction was a
preferred direction for ancient Egyptians who observed motion of heavenly
bodies regularly. It was direction toward the fixed North star. They did not
connect direction of the river flow with the plane slope, as we do now. They
connected the direction of the river flow with the preferred spatial
direction towards the North. We are interested now what kind of mistake made
ancient Egyptians, believing that all rivers flow towards the North, and how
could they to overcome their delusion.

Their delusion was not a logical mistake, because the logic has no relation
to this mistake. The delusion was connected with associative property of
human thinking, when the property $P$ is attributed to the object $O$ on the
basis that in all known cases the property $P$ accompanies the object $O$.
Such an association may be correct or not. If it is erroneous, as in the
given case, it is very difficult to discover the mistake. At any rate it is
difficult to discover the mistake by means of logic, because such
associations appear before the logical analysis, and the subsequent logical
analysis is carried out on the basis of the existing associations. Let us
imagine that in the course of a travel an ancient Egyptian scientist arrived
the Tigris, which is the nearest to Egypt river. He discovers a water stream
which flows, first, not directly and, second, not towards the North. Does he
discover his delusion? Most likely not. At any rate not at once. He starts
to think that the water stream, flowing before him, is not a river. A ground
for such a conclusion is his primordial belief that ''real'' river is to
flow, first, directly and, second, towards the North. Besides, the Nile was
very important in the life of ancient Egyptians, and they were often apt to
idolize the Nile. The delusion about direction of the river flow could be
overcame only after that, when one has discovered sufficiently many
different rivers, flowing towards different directions, and the proper
analysis of this circumstance has been carried out.

Thus, to overcome the associating delusion, it is not sufficient to present
another object $O$, which has not the property $P$, because one may doubt of
whether the presented object is to be classified really as the object $O$.
Another attendant circumstances are also possible.

If the established association between the object and its property is
erroneous, one can speak on associative delusion or on associative
prejudice. The usual method of the associative delusions overcoming is a
consideration of a wider set of phenomena, where the established association
between the property $P$ and the object $O$ may appear to be violated, and
the associative delusion is discovered.

In this paper the associative delusions in natural sciences, mainly in
physics are discussed. The associative delusions (AD) are very stable. If
they have been established, they are overcame very difficultly, because they
cannot be disproved logically. But there is an additional complication. The
usual mistake is overcame easily by the scientific community, as soon as it
has been overcame by one of its members. The corresponding article is
published, and the scientific community takes it into account, and the
mistake is considered to be corrected.

A different situation arises with the associative delusions (AD). Discovery
of the associative delusion (AD), and publication of corresponding article
do not lead to acknowledgment of AD as a delusion or mistake. The scientific
community continue to insist on the statement, that the considered in the
article AD is not a mistake in reality, and that the author of this paper
himself makes a mistake. A long controversy arises. Sometimes it leads to a
conflict, as in the case of conflict between the Ptolemaic doctrine and that
of Copernicus. Finally, the truth celebrates victory, but the way to this
victory appears to be long and difficult.

Apparently, the reason of the AD stability lies in obviousness and
habitualness of those statements, which appear to be associative delusions
afterwards. On the ground of these statements one constructs scientific
conceptions, which agree with experimental data and observations. Declaring
these habitual statements to be a delusions, one destroys existing
scientific conceptions and tries to construct new conceptions. It is always
very difficult for the scientific community.

In the science history a series of associative delusions is known. Let us
list them in the chronological order.

AD.1. The antipodes paradox, generated by that the gravitational field
direction is connected with a preferred direction in the space, but not with
the direction towards the Earth centre.

AD.2. The Ptolemaic doctrine in the celestial mechanics, where the property
of being the "universe" centre was attributed to the Earth, whereas the Sun
is such a centre.

AD.3. Prejudices against the Riemannian geometry in the second half of the
XIX century are connected with that the Cartesian coordinate system was
considered to be an attribute of any geometry, whereas it was only a method
of the Euclidean geometry description.

AD.4. Impossibility of employment of the pure metrical conception of
geometry, connected with the associative delusion, that the concept of the
curve is considered to be a fundamental concept of any geometry, whereas the
curve is only a geometrical object, used in the Euclidean and Riemannian
geometry.

AD.5. Impossibility of construction of dynamical conception of statistical
description (DCSD), connected with the associative delusion, that any
statistical description is considered to be produced in terms of the
probability theory, and the probability concept is a fundamental concept of
any statistical description.

AD.6. Identification of individual particle ${\cal S}$ with the
statistically averaged particle $\langle {\cal S}\rangle $, used at the
conventional interpretation of quantum mechanics. Such an identification is
a kind of associative delusion, when the individual particle properties $%
{\cal S}$ are attributed to the statistically averaged particle $\langle 
{\cal S}\rangle $ and vice versa. The Schr\"{o}dinger cat paradox and some
other quantum mechanics paradoxes, connected with the wave function
reduction, are corollaries of this identification.

AD.7. The forced identification of energy and Hamiltonian, used in
relativistic quantum field theory (QFT), is also an associative delusion. As
any associative delusion this identification is connected with attributing
properties of one object to another one. In the given case the coincidence
of energy and Hamiltonian for a free nonrelativistic particle is considered
to be a fundamental property of any particle without sufficient foundations.
We shall denote this delusion symbolically by means of $E=H$, where $E$ is
the energy, and $H$ is the Hamiltonian.

The first three of the seven listed delusions (AD.1 -- AD.3) had been
overcame to the beginning of XX century, though a detailed analysis of these
overcoming is, maybe, absent in the literature. As to AD.4 -- AD.7, the
scientific community is yet destined to overcome them. Besides these ADs
exist simultaneously, and the order of their listing corresponds basically
to their importance rather than to chronology.

The purely metric conception of geometry (CG), where all information on
geometry is given by means a distance between to space points, is the most
general conception of geometry (CG). It generates the most complete list of
geometries, suitable for the space-time description. AD.4 discriminates the
purely metric CG. As a result instead of it one uses Riemannian CG,
generating incomplete list of possible geometries. The true space-time
geometry is absent in this list, and we are doomed to use the Minkowski
geometry for the space-time description. The Minkowski geometry is incorrect
geometry for small space-time scales, i.e. in the micro world. In the true
space-time geometry the micro particle motion is primordially stochastic,
and the properties of the geometry are an origin of this stochasticity. In
the Minkowski geometry the motion of any particle is deterministic, and
incorrectness of the Minkowski geometry lies in this fact.

AD.5 leads to impossibility of a construction of a consecutive statistical
description of the stochastically moving micro particles (electrons,
positrons, etc.) , although it is doubtless that quantum mechanics,
describing the regular component of this motion, is a statistical theory.
AD.4 and AD.5 establish such a situation, when one is forced to use a series
of additional hypotheses (quantum mechanics principles) for a correct
description of observed quantum phenomena. It is much as the Ptolemeus used
a series of additional construction (epicycles, differents) for explanation
of observed motion of heavenly bodies. They were needed for compensation of
AD.2.

Overcoming of AD.4 and AD.5 admits one to eliminate the quantum mechanics
principles and to construct the quantum phenomena theory as a consecutive
statistical description of stochastic micro particle motion. At such a
description the micro particle stochasticity has a geometric origin, i.e. it
is generated by the space-time geometry. The consecutive statistical
description of the micro particles appears as a result of a construction of
the dynamical conception of statistical description (DCSD), that becomes to
be possible after overcoming of AD.5. The elimination of the quantum
mechanics principles after overcoming of AD.4 and AD.5 resembles the
eliminating of Ptolemaic epicycles, when they stopped to be necessary after
overcoming AD.2 and the subsequent transition from the Ptolemaic doctrine to
that of Copernicus.

AD.6 has not such a global character as AD.4 and AD.5. It concerns mainly
the interpretation of the measurement concept in quantum mechanics.

AD.7 has not the global character also. It acts only in the scope of the
relativistic quantum field theory (QFT). QFT in itself is only a section of
the Ptolemaic conception, i.e. a conception, which uses additional
hypotheses (quantum mechanics principles), compensating incorrect choice of
the space-time model. AD.7 (identification of energy and Hamiltonian $E=H$)
generates a series of difficulties in QFT (non-stationary vacuum, necessity
of the perturbation theory and some other). In fact, there is no necessity
the energy -- Hamiltonian identification $E=H$. The secondary quantization
can be carried out without imposing this constraint \cite{R72,RY001}. The
condition $E=H$ appears to be inconsistent with dynamic equations.
Imposition of this constraint makes QFT to be inconsistent. On one hand,
such an inconsistency leads to above mentioned difficulties, but on the
other hand, such an inconsistency admits one to explain the pair production
effect, because any inconsistent theory admits one to explain all what you
want. One needs only to show sufficient ingenuity. On one hand, elimination
of the constraint $E=H$ leads to a theory which is consequent in the scope
of quantum theory and free from the above mentioned difficulties, but on the
other hand, it leads to that the theory stops to describe the pair
production effect. This deplorable fact means only, that the undertaken
attempt of the FTP construction on the basis of unification of the
relativity principles with those of quantum mechanics failed, and one should
search for alternative conception.

Let us take into account that the quantum mechanics is a compensating
(Ptolemaic) conception, i.e. just as the quantum mechanics principles have
been invented for compensation of AD.4 and AD.5, as the Ptolemaic epicycles
have been invented for compensation of AD.2. Then an attempt of unification
of quantum mechanics principles with the relativity ones is as useless, as
an attempt of introduction of Ptolemaic epicycles in Newtonian mechanics.

Apparently, the conception, appeared after overcoming of AD.4 and AD.5, is a
reasonable alternative to QFT. Such a conception is consistently
relativistic and quantum (in the sense that it contains the quantum constant 
$\hbar $, contained explicitly in the space-time metric). It does not
contain the quantum mechanics principles, and one does not need to unite
them with the relativity principles. We shall refer to this conception as
the model conception of quantum phenomena, distinguishing it from
conventional quantum mechanics, which will be referred to as axiomatic
conception of quantum phenomena. The difference between axiomatic conception
and model one is much as the difference between the thermodynamics and the
statistical physics. The thermodynamics may be qualified as the axiomatic
conception of thermal phenomena, whereas the statistical physics may be
qualified as the model conception of thermal phenomena. The transition from
the axiomatic conception to the model one was carried out after a
construction of the ''calorific fluid'' model (chaotic motion of molecules),
and the thermodynamics axioms, describing properties of the fundamental
thermodynamical object -- ''calorific fluid'', stopped to be necessary.
Concept of ''calorific fluid'' is not used usually in the statistical
physics, but if it is introduced, its properties are determined from its
model (chaotic molecular motion).

Similar situation takes place in the interrelations between the axiomatic
and model conceptions of quantum phenomena. In the axiomatic conception
there is a fundamental object, called the wave function. Its properties are
determined by the quantum mechanics principles. The wave function is that
object, which differs the quantum mechanics from the classical one, where
the wave function is absent. In the model conception one constructs a
''model of the wave function'' \cite{R99}. Thereafter the wave function
properties are obtained from this model, and one does not need the quantum
mechanics principles. Axiomatic and model conceptions lead to the same
result in the nonrelativistic case, but in the relativistic case the results
are different, in general. For instance, application of the model conception
to investigation of the dynamic system ${\cal S}_{{\rm D}}$, described by
the Dirac equation, leads to another result \cite{R001}, than investigation,
produced by conventional methods in the scope of the axiomatic conception.
In the first case the classical analog of the Dirac particle ${\cal S}_{{\rm %
D}}$ is a relativistic rotator, consisting of two charged particles,
rotating around their common center of mass. In the second case the
classical analog is a pointlike particle, having spin and magnetic moment.

An existence of the associative delusion does not permit one to construct a
rigorous scientific conception. The constructed building appears to be a
compensating (Ptolemaic) conception, where an incorrect statement is
compensated by means of additional suppositions. In general, the Ptolemaic
conception is not true. But there are such fields of its application, where
its employment leads to correct results, which agree with observations and
experimental data. For instance, in the scope of the Ptolemaic doctrine one
can choose such epicycles and differents for any planet, that one can
calculate its motion in a sufficient long time so, that predictions agree
with observations. But there is a class of the celestial mechanics problems,
which could not be solved in the scope of the Ptolemaic doctrine. For
instance, in the scope of this doctrine one cannot solve such a problem:
when and with what velocity should one throw a stone from the Earth's
surface, in order that it could drop on the Moon. In the scope of the
Ptolemaic doctrine one cannot discover the gravitation law and construct the
Newtonian mechanics. The associative delusion, embedded in the ground of the
Ptolemaic doctrine and disguised by means of compensating hypotheses,
hindered the progress of celestial mechanics. As far as in that time the
celestial mechanics was the only exact natural science, AD hindered the
normal development of natural sciences at all. The development of natural
sciences went to blind alley. After overcoming of AD.2 the natural sciences
development was accelerated strongly.

The same situation takes place with the quantum mechanics. Although at the
first acquaintance the quantum mechanics seems to be a disordered collection
of rules for calculation of mathematical expectations, nevertheless, in the
nonrelativistic case an employment of these rules leads to results which
agree with experiments. Accepting the quantum mechanical principles, the
nonrelativistic quantum mechanics as a whole is a consistent conception,
which describes excellently a wide class of physical phenomena. But at the
transition to the field of relativistic phenomena (pair production,
elementary particles theory) the quantum principles stop to be sufficient.
One is forced to introduce new suppositions. The further the quantum theory
advances in the field of relativistic phenomena, the more new suppositions
are to be introduced for descriptions of observed phenomena. This is an
indirect indication, that the conventional way of the quantum theory
development comes to a blind alley.

Investigation of possible methods of the associative delusions overcoming is
a subject of this paper. On one hand, overcoming of any special associative
delusion needs a knowledge of the subject of investigation and a
professional approach to the investigation of the phenomenon. On the other
hand, the Ptolemaic conceptions have some common properties, and a work with
them has some specific character, which should be known, if we want to
overcome corresponding ADs effectively.

First, it is very difficult to discover the associative delusion. Indirect
indications of AD are an increasing complexity of the theory and a necessity
of new additional suppositions. These indications show that the associative
delusion does exist, but they do not permit one to determine, what is this
AD.

Second, the work with Ptolemaic conceptions, i.e. with conceptions,
containing AD, generates a special pragmatic style (P-style) of
investigations. the P-style lies in the fact that one searches all possible
ways of explanation and calculation of the considered phenomenon. Of course,
different versions, considered at such an approach, are restricted by the
existing mathematical technique and by the possibilities of the researcher's
imagination. But these restrictions are essentially slighter, than the
restrictions imposed by the classical style (C-style) of investigations. The
classical style (C-style) is the style of investigations, fully developed in
the natural sciences to the end of the XIX century.

Unprejudiced reader will agree that the delusions AD.1 -- AD.3, having been
overcame, are delusions indeed, and that it was worth to overcame them. But
it is rather doubtless that he agrees at once that AD.4 -- AD.7 are also
delusions and that they are to be overcame. If it were so, then AD.4 -- AD.7
have been overcame many years ago. Of course, ADs are undesirable as any
other delusions. One should eliminate them, if it is possible. But one
should not consider them as misunderstandings, or manifestations of
researcher's stupidity. ADs are inevitable attributes of the cognitive
processes, as far as they are conditioned by the restriction of the field of
investigated phenomena at the initial stage of investigations. ADs were in
the past, they exist now, and apparently, they will exist in the future. We
should know, how to live with them and to make investigation. The situation
resembles the situation with a noise. We transmit information at presence of
a noise, and we know that the noise is undesirable, that the noise should be
removed, and that, unfortunately, it cannot be removed completely.

One should study associative delusions, their properties and the influence
on the style of thinking and investigations of researchers, which are forced
to work under conditions of the associative delusions presence.
Investigation of ADs properties and possibilities of their overcoming is a
goal of this paper. We begin with detailed investigations of AD.4 -- AD.7,
to make sure that they are delusions indeed and to understand how to
overcome them. It is very important, because experience of overcoming of
AD.2 (Ptolemaic doctrine) shows that the overcoming process is very
difficult for scientific community.

Usually these difficulties are connected with a negative role of the
Catholic church. B.V.~Raushenbach \cite{R2001} considers that the position
of the Catholic church is not the case. It was incompetent in problems of
celestial mechanics. It agreed simply with opinion of the most of that time
researchers. Most of scientists of that time were priests, and
B.V.~Raushenbach considers that they used the Catholic church simply as a
tool for a fight against proponents of the Copernicus doctrine. Experience
of the author in attempts of overcoming of AD.4 -- AD.7 shows, that this is
B.V.~Raushenbach, who is right.

In sections 2 -- 5 one considers properties of AD.4 -- AD.7. In the sixth
section influence of associative delusions on the style of investigations is
considered. In sections from seventh to twelfth the details of history of
the AD.4 -- AD.7 overcoming are presented in the form, as the author of this
paper saw it.

\section{Conception of geometry and a correct choice of the space-time
geometry}

The conception of geometry (CG) is considered to be the method (a set of
rules), by means of which the geometry is constructed. The proper Euclidean%
\footnote{%
We use the term ''Euclidean geometry'' as a collective concept with respect
to terms ''proper Euclidean geometry'' and ''pseudoeuclidean geometry''. In
the first case the eigenvalues of the metric tensor matrix have similar
signs, in the second case they have different signs.} geometry can be
constructed on the basis of different geometric conceptions.

For instance, one can use the Euclidean axiomatic conception (Euclidean
axioms), or the Riemannian conception of geometry (dimension, manifold,
metric tensor, curve). One can use the topology-metric conception of
geometry (topological space, metric, curve). In any case one obtains the
same proper Euclidean geometry. From point of view of this geometry it is of
no importance which of possible geometric conceptions is used for the
geometry construction.

But if we are going to choose a geometry for the real space-time, it is very
important, that the list of all possible geometries, suitable for the
space-time description, would be complete. If the true space-time geometry
is absent in this list, we are doomed to a choice of a false geometry
independently of the method which is used for a choice of the space-time
geometry. Thus, a determination of the complete list of all possible
geometries is a necessary condition of a correct choice of the space-time
geometry. In turn the determination of the possible geometries list depends
on the conception of geometry (CG), which is used for determination of the
list of possible geometries. Any of possible CG contains information of two
sorts: (1) non-numerical information in the form of concepts, axioms and
propositions, formulated verbally, (2) numerical information in the form of
numbers and numerical functions of space points. In different CG this
information is presented differently. \bigskip

\noindent 
\centerline{$
\begin{array}{|c|c|c|}
\hline
& & \\
\text{\it title of CG} &
\begin{array}{c}
\text{\it non-numerical}  \\
\text{\it information}
\end{array}
&
\text{\it numerical information}
\\ & &
\\
\hline
\text{Euclidean CG} & \text{Euclidean axioms} & \emptyset
\\
\hline
\text{Riemannian CG}\label{tab} &
\begin{array}{c}
\text{Manifold, curve} \\
\text{coordinate system}
\end{array}
& n,\;\;g_{ik}\left( x\right)
\\
\hline
\begin{array}{c}
\text{topology-} \\ \text{metric CG}
\end{array}
&
\begin{array}{c}
\text{topological space,} \\
\text{curve}
\end{array}
&
\begin{array}{c}
\rho \left( P,Q\right) \geq 0,\\
\rho \left( P,Q\right) =0,\text{\ \ iff \
\thinspace }P=Q \\
\rho \left( P,Q\right) +\rho \left( Q,R\right) \geq \rho \left( P,R\right)
\end{array}  \\
\hline
\begin{array}{c}
\text{purely} \\ \text{metric CG}
\end{array}
& \emptyset  & \sigma \left( P,Q\right) =\frac{1}{2}\rho ^{2}
\left( P,Q\right) \in {\Bbb R} \\
\hline
\end{array}
$} \bigskip

Varying the numerical information at fixed the non-numerical one, we obtain
different geometries in the scope of the same conception of geometry.
Varying continuously numbers and functions, constituting numerical
information of CG, one obtains a continuous set of geometries, each of them
differs slightly from the narrow one. Any admissible value of numerical
information is attributed some geometry in the scope of the given CG. One
can also change non-numerical information, replacing one axiom by another.
But at such a replacement the geometry changes step-wise, and one should
monitor that replacements of one axiom by another does not lead to
inconsistencies. It is complicated and inconvenient. It is easier to obtain
new geometries in the scope of the same conception, changing only numerical
information.

One can see from this table, that different CG have different capacity of
the numerical information and generate the geometry classes of different
power. The Euclidean CG does not contain the numerical information at all.
Vice versa, the purely metric CG contains only numerical information and
generates the most powerful class of geometries which will be referred to as
tubular geometries (or briefly T-geometries).

The T-geometry has many attractive features. Firstly, it is very simple and
realizes the simple attractive idea, that for determination of a geometry on
a set $\Omega $ of points $P$ it is sufficient to give the distance $\rho
(P,Q)$ between all pairs $\left\{ P,Q\right\} $ of points of the set $\Omega 
$. In fact, the distance $\rho (P,Q)$ is determined by means of the world
function $\sigma =\frac{1}{2}\rho ^{2}$ on the set $\Omega \times \Omega $.
In spite of simplicity and attractiveness of this idea the existence
possibility in itself of the purely metric CG was being problematic for a
long time. K.~Menger \cite{M28} and J.L.~Blumenthal \cite{B53} tried to
construct so called distance geometry, which was founded on the concept of
distance in a larger degree, than it is made in the topology-metric CG. But
they failed to construct the purely metric CG. The reason of the failure was
AD.4. The statement of necessary and sufficient conditions of the geometry
Euclideness in terms of the world function $\sigma $, given on the set $%
\Omega \times \Omega $, was a crucial step in construction of the purely
metric CG. The prove \cite{R90,R01,Ra01} of the fact, that the Euclidean
geometry can be constructed in terms of only $\sigma $ meant a possibility
of construction of any T-geometry in terms of $\sigma $. It meant existence
of the purely metric CG.

In the scope of purely metric CG all information on geometry is derived from
the world function. In particular, if one can introduce a dimension of the
space $\{\Omega ,\sigma \}$, this information can be derived from the world
function. From the world function one can derive information on continuity,
or discontinuity of the space $\{\Omega ,\sigma \}$. In the case of
continuous geometry the information on the coordinate systems and metric
tensor can be also derived from the world function. In T-geometry there is
an absolute parallelism (which is absent in Riemannian geometries). Besides
the T-geometry has a new property -- nondegeneracy.

The geometry is called a nondegenerate, if there are many vectors $%
\overrightarrow{P_{0}Q}$ of fixed length $|\overrightarrow{P_{0}Q}|$,
parallel to the vector $\overrightarrow{P_{0}P_{1}}$. In the degenerate
geometry there is only one such a vector $\overrightarrow{P_{0}Q}$.

Nondegeneracy of T-geometry may be conceived as follows. Any T-geometry can
be obtained from the Euclidean geometry by means of its deformation (i.e. a
change of distance $\rho (P,Q)$ between the space points). At such a
deformation the geometrical objects of Euclidean geometry change their
shape. If the obtained T-geometry is degenerate, the Euclidean straights
transform to lines, which are curved lines, in general. But it is possible
such a deformation, that the straight of $n$-dimensional Euclidean space
converts into $(n-1)$-dimensional tube. For it would be a possible, the
straight is to be defined as a set of points, possessing some property of
the Euclidean straight. Definition of the straight as a curve, possessing
some property of the Euclidean straight prohibits automatically deformation
of the Euclidean straight into $(n-1)$-dimensional tube and discriminates
nondegenerate geometries.

If one considers nondegenerate T-geometry of the space-time, the motion of
free particles in such a space-time appears to be stochastic, although the
geometry in itself (i.e. the world function $\sigma $) is deterministic. In
other words, nondegeneracy of the space-time geometry generates an
indeterminism.

In the Riemannian CG the deformation, converting a line into a tube, is
forbidden. It is connected with AD.4, according to which the curve is a
fundamental object of geometry, and there are not to be such geometries,
where the curve would be replaced by a surface. It is in this point, where
AD.4 discriminates purely metric CG and T-geometries, generated by this CG.
As a corollary the list of possible geometries reduces strongly. The true
space-time geometry fall out of the list of possible geometries, and one
chooses a false model for the space-time.

In the present time one uses the Riemannian conception for obtaining the
space-time geometry. In the simplest case, when one can neglect gravitation,
the space-time is uniform, isotropic and flat. In the scope of the
Riemannian geometry there is only one flat uniform isotropic geometry. It is
the Minkowski geometry, for which the world function has the form: 
\begin{equation}
\sigma _{{\rm M}}\left( x,x^{\prime }\right) =\sigma _{{\rm M}}\left( t,{\bf %
x},t^{\prime },{\bf x}^{\prime }\right) =\frac{1}{2}\left( c^{2}\left(
t-t^{\prime }\right) ^{2}-\left( {\bf x-x}^{\prime }\right) ^{2}\right)
\label{a1.1}
\end{equation}
where $c$ is the speed of the light, and $x=\left\{ t,{\bf x}\right\} $, \ $%
x^{\prime }=\left\{ t^{\prime },{\bf x}^{\prime }\right\} $ are coordinates
of two arbitrary points in the space-time.

Thus, in the case of Riemannian CG the problem of choosing space-time
geometry does not appear. It is determined uniquely.

The topology-metric CG cannot be applied to the space-time, because it
supposes that $\sigma \left( P,Q\right) =\frac{1}{2}\rho ^{2}\left(
P,Q\right) \geq 0$, whereas in the space-time there are spacelike intervals,
for which $\sigma \left( P,Q\right) <0$.

The purely metric CG generates a whole class of flat uniform isotropic
T-geometries, labelled by a function of one argument. In this case the world
function has the form 
\begin{equation}
\sigma \left( x,x^{\prime }\right) =\sigma _{{\rm M}}\left( x,x^{\prime
}\right) +D\left( \sigma _{{\rm M}}\left( x,x^{\prime }\right) \right) ,
\label{a1.2}
\end{equation}
where $\sigma _{{\rm M}}$ is the world function for the Minkowski space (\ref
{a1.1}), and the function $D$ is an arbitrary function, labelling possible
flat uniform isotropic geometries. These geometries differ one from another
in the shape of tubes, obtained as a result of the Euclidean straight
deformation. Hence, they differ in the stochasticity character of the free
particles motion. For the purely metric CG the problem of choice of the
space-time geometry is very important, because there are many uniform
isotropic geometries. To set $D\equiv 0$ in (\ref{a1.2}) and choose the
Minkowski geometry would be incorrect, because in the Minkowski geometry the
motion of particles is deterministic. But it is well known that the motion
of real micro particles (electrons, positrons, etc.) is stochastic. In other
words, experiments with single particles are irreproducible. Only
distributions of results, i.e. results of mass experiments with many
similarly prepared particles are reproducible. These distributions of
results are described by quantum mechanics, constructed on the basis of some
additional hypotheses, known as principles of quantum mechanics.

When there are such space-time geometries, where the motion of particles is
primordially stochastic, one cannot consider as reasonable such an approach,
where at first one chooses the Minkowski geometry with deterministic motion
of particles, and thereafter one introduces additional suppositions (quantum
mechanics principles), providing a description of the stochastic motion of
micro particles. It would be more correct to choose the space-time geometry
in such a way, that a statistical description of stochastic motion of micro
particles would describe correctly experimental data. As far as the quantum
mechanics describes all nonrelativistic experiments very well, it is
sufficient to choose the space-time so, that the statistical description of
stochastic motion of micro particles would agree with predictions of quantum
mechanics.

At first sight, it seems that the quantum effects cannot be explained by
peculiarities of geometry, because intensity of quantum effects depends on
the particle mass essentially, and the mass is such a characteristic of a
particle, which is not connected with a geometry. It seems that influence of
a geometry on the particle motion is to be similar for particles of any
mass. In reality the influence of geometry does not depend on particle
motion only in the nondegenerate geometry (Minkowski geometry). In the
space-time with the nondegenerate geometry the particle mass, as well as its
momentum are geometrical characteristics.

The world tube of the particle with the mass $m$ is described by the broken
world tube ${\cal T}_{{\rm br}}$, which is determined by a sequence of the
break points $\{P_{i}\}$, \thinspace \thinspace \thinspace\ $i=0,\pm 1,\pm
2,\ldots $. The adjacent points $P_{i}$, $P_{i+1}$ are connected between
themselves by a segment ${\cal T}_{[P_{i}P_{i+1}]}$ of the straight. This
segment is determined by the relation 
\begin{equation}
{\cal T}_{[P_{i}P_{i+1}]}=\left\{ R|S\left( P_{i},R\right) +S\left(
R,P_{i+1}\right) =S\left( P_{i},P_{i+1}\right) \right\}  \label{a1.3}
\end{equation}
where $S\left( P_{i},P_{i+1}\right) =\sqrt{2\sigma \left(
P_{i},P_{i+1}\right) }$ is the distance between the points $P_{i}$ and $%
P_{i+1}$. The set of points $\{P_{i}\}$, \thinspace \thinspace \thinspace\ $%
i=0,\pm 1,\pm 2,\ldots $ will be referred to as the skeleton of the tube $%
{\cal T}_{{\rm br}}$.

In the proper Euclidean geometry as well as in the Minkowski geometry (for
timelike interval $S^{2}\left( P_{i},P_{i+1}\right) >0$) the set of points (%
\ref{a1.3}) forms a segment of the straight line, connecting points $P_{i}$, 
$P_{i+1}$. In the nondegenerate geometry the set ${\cal T}_{[P_{i}P_{i+1}]}$
forms a three-dimensional cigar-shaped surface with the ends at the points $%
P_{i},$ $P_{i+1}$.

The vector $\overrightarrow{P_{i}P_{i+1}}=\left\{ P_{i},P_{i+1}\right\} $ is
interpreted as the particle 4-momentum on the segment ${\cal T}%
_{[P_{i}P_{i+1}]}$ of the particle world tube ${\cal T}_{{\rm br}}$ 
\begin{equation}
{\cal T}_{{\rm br}}=\bigcup\limits_{i}{\cal T}_{[P_{i}P_{i+1}]}.
\label{a1.4}
\end{equation}
The length $\left| \overrightarrow{P_{i}P_{i+1}}\right| =S\left(
P_{i},P_{i+1}\right) =\sqrt{2\sigma \left( P_{i},P_{i+1}\right) }$ of the
vector $\overrightarrow{P_{i}P_{i+1}}$ is the geometrical mass $\mu $ of the
particle, expressed in units of length. The universal constant $b$ connects
the geometrical mass $\mu $ with the usual mass $m$ of the particle. 
\begin{equation}
m=b\mu =bS\left( P_{i},P_{i+1}\right) ,\qquad i=0,\pm 1,\pm 2,...\qquad
\lbrack b]=\text{g/cm}  \label{a1.5}
\end{equation}

All segments ${\cal T}_{[P_{i}P_{i+1}]}$,\ $i=0,\pm 1,\pm 2,...$ has the
same length $\mu =m/b$. Thus, in general, $m$ is a geometrical
characteristic of the particle, but in the case of the Minkowski geometry
one cannot determine the particle mass, using the world line shape, because
one cannot determine points $P_{i}$ of the world line ${\cal T}_{{\rm br}}$
skeleton on the basis of the world line shape. In the case of nondegenerate
space-time geometry the points $P_{i}$ of the skeleton are end points of the
cigar-shaped segments ${\cal T}_{[P_{i}P_{i+1}]}$. They can be determined
via intakes of the broken tube (\ref{a1.4}). Interval $S\left(
P_{i},P_{i+1}\right) $ between adjacent points $P_{i}$ of the skeleton
determines the geometrical mass $\mu $ of the particle.

For a free particle the 4-momenta $\overrightarrow{P_{i}P_{i+1}}$ and $%
\overrightarrow{P_{i+1}P_{i+2}}$ of two adjacent segments ${\cal T}%
_{[P_{i}P_{i+1}]}$ and ${\cal T}_{[P_{i+1}P_{i+2}]}$ are parallel $%
\overrightarrow{P_{i}P_{i+1}}\uparrow \uparrow \overrightarrow{P_{i+1}P_{i+2}%
}$. In the Minkowski geometry there is only one vector $\overrightarrow{%
P_{i+1}P_{i+2}}$ of the length $\mu $, parallel to timelike vector $%
\overrightarrow{P_{i}P_{i+1}}$. Hence, if the vector $\overrightarrow{%
P_{0}P_{1}}$ is fixed, all other vectors $\overrightarrow{P_{i}P_{i+1}}$ $%
\;\;i=1,2,...$ are determined uniquely. In other words, in the Minkowski
geometry the total world line ${\cal T}_{{\rm br}}$ is determined uniquely,
provided one of its segments is fixed. It means that the motion of a free
particle in the space-time with Minkowski geometry is deterministic.

In the space-time with nondegenerate geometry there are many vectors $%
\overrightarrow{P_{i+1}P_{i+2}}$ of the length $\mu $, parallel to the
timelike vector $\overrightarrow{P_{i}P_{i+1}}$. It means that the end $%
P_{i+2}$ of the vector $\overrightarrow{P_{i+1}P_{i+2}}$ is not determined
uniquely, even if the vector $\overrightarrow{P_{i}P_{i+1}}$ is fixed. Other
points $P_{i+3}$, $P_{i+4}$,... are not determined uniquely also. It means
that the broken tube ${\cal T}_{{\rm br}}$ is stochastic. Thus, the motion
of a free particle in the space-time with nondegenerate geometry is
stochastic. The character and intensity of the stochasticity depend on the
form of the function $D\left( \sigma _{{\rm M}}\right) $ in the relation (%
\ref{a1.2}).

Supposing that the statistical description of stochastic world tubes gives
the same result, as the quantum-mechanical description in terms of the
Schr\"{o}dinger equation, one can calculate the distortion function $D\left(
\sigma _{{\rm M}}\right) $. The calculation gives \cite{R91} 
\begin{eqnarray}
D &=&D\left( \sigma _{{\rm M}}\right) =\left\{ 
\begin{array}{ccc}
d & \text{if} & \sigma _{{\rm M}}>\sigma _{0} \\ 
0 & \text{if} & \sigma _{{\rm M}}\leq 0
\end{array}
\right.  \label{a1.6} \\
d &=&\frac{\hbar }{2bc}=\text{const}\approx 10^{-21}\text{cm,\qquad }\sigma
_{0}=\text{const}\approx d  \nonumber
\end{eqnarray}
Here $\hbar $ is the quantum constant, and $b\approx 10^{-17}$g/cm is a new
universal constant. Inside the interval $(0,\sigma _{0})$ values of the
function $D\left( \sigma _{{\rm M}}\right) $ are not yet determined.

From the three-dimensional viewpoint the micro particle is a pulsating
sphere. Period $T$ of pulsations depends on the particle mass $m$. It is
determined by the relation $T=m/(bc)$, where $b$ is the universal constant.
The maximal sphere radius $R_{{\rm max}}\approx \sqrt{d}$ does not depend on
the particle mass. Approximately one can assume that in the period $T$ the
sphere radius increases from zero up to maximal value $R_{{\rm max}}$, and
then it reduces to zero. In the period $T$ the sphere centre moves along the
straight line uniformly. At the collapse moment a random jump-like change of
velocity takes place. In the coordinate system, where the sphere is at rest
the velocity jump is equal approximately to $R_{{\rm max}}/T\approx
m^{-1}(\hbar bc/2)^{1/2}$. The lesser is the particle mass the larger is the
velocity jump. Besides, the period $T$ depends on the particle mass. As a
result for the particle of small mass the random velocity jumps are happens
more often and have the larger magnitude. Thus, choosing the space-time
geometry in the form (\ref{a1.2}), (\ref{a1.6}), one can explain all
nonrelativistic quantum effects without referring to quantum principles.
Such a space-time geometry is more correct, than the Minkowski geometry,
because in this case one does not need additional hypotheses in the form of
quantum principles. In such a geometry the quantum constant appears in the
theory together with the distortion function (\ref{a1.6}). It is an
attribute of the space-time, that agrees with the universal character of the
quantum constant $\hbar $.\label{01}

\section{Dynamical conception of statistical description}

As we have mentioned, the choice of the space-time geometry is determined by
the condition that the statistical description of the stochastic motion of
particles is to coincide with the norelativistic quantum-mechanical
description. It means that the quantum mechanics is to be represented as a
statistical description of randomly moving particles. In the end of XIX
century the thermodynamics was presented as a statistical description of
chaotically moving molecules. After this representation many researchers
thought that something like that can be made with the quantum mechanics. It
is a common practice to think that any statistical description is produced
in terms of the probability theory. In this point we meet AD.5, where it is
supposed that there is no statistical description without the probability
theory. Attempts \cite{M49,F52} of formulating the quantum mechanics in
terms of the probability theory failed. The fact is that, attempting to
represent the quantum mechanics as a statistical description of stochastic
particle motion, one overlooks usually, that the random component of the
particle motion can be relativistic, whereas the regular component remains
to be nonrelativistic.

The probability theory, applied successfully to the statistical physics for
statistical description of the chaotic molecule motion, is not suitable for
a description of the stochastic motion of relativistic particles. The fact
is that, the employment of the probability density supposes splitting of all
possible system states into sets of simultaneous independent events. In the
relativistic theory it cannot be made for a continuous dynamic system, as
far as there is no absolute simultaneity in the special relativity. The
simultaneity at some coordinate system cannot be used also, because the
coordinate system is a method of description. Application of the probability
theory and of the conditional simultaneity (simultaneity at some coordinate
system) means an application of the statistics to the description methods
instead of the necessary calculation of the dynamic system states.

One can overcome the appeared obstacle, rejecting employment of the
probability theory at the statistical description. Indeed, the term
''statistical description'' means only that one considers many identical, or
almost identical objects. Application of the probability theory in the
statistical description is not necessary, because it imposes some
constraints on the method of the description, that is undesirable. For
instance, the probability density must be nonnegative, and sometimes this
constraint cannot be satisfied.

In the nonrelativistic physics the physical object to be statistically
described is a particle, i.e. a point in the usual space or in the phase
one. The density of points (particles) in the space is nonnegative, it is a
ground for introduction of the probability density concept. In the
relativistic theory the physical object to be statistically described is a
world-line in the space-time. The density of world lines in the vicinity of
some point $x$ is a 4-vector, which cannot be a ground for introduction of
the probability density. The alternative version, when any world line is
considered to be a point in some space ${\cal V}$, admits one to introduce
the concept of the probability density in the space ${\cal V}$ of world
lines. But such a description is non-local, as far as two world lines,
coinciding everywhere except for some remote regions, are represented by
different points in ${\cal V}$, and this points are not close, in general.
In other words, such an introduction of the probability is very inconvenient.

To get out of this situation, one needs to reject from employment of the
probability theory at the statistical description. Instead of the
probabilistic conception the dynamical conception of statistical description
(DCSD) should be used. Instead of the stochastic system ${\cal S}_{{\rm st}}$%
, for which there are no dynamic equations, one should use a set ${\cal E}[N,%
{\cal S}_{{\rm st}}]$, consisting of large number $N$ of identical
independent systems ${\cal S}_{{\rm st}}$ and known as the statistical
ensemble of systems ${\cal S}_{{\rm st}}$. The statistical ensemble ${\cal E}%
[N,{\cal S}_{{\rm st}}]$ forms a deterministic dynamical system, for which
there are dynamic equations, although they do not exist for elements ${\cal S%
}_{{\rm st}}$ of the statistical ensemble. The statistical description lies
in the fact that one investigates properties of ${\cal E}[N,{\cal S}_{{\rm st%
}}]$ as a deterministic dynamic system, and on the basis of this
investigation one makes some conclusions on properties of its elements
(stochastic systems ${\cal S}_{{\rm st}}$)). As far as one investigates a
dynamic system (statistical ensemble) and its properties, there is no
necessity to use the concept of probability.

Along with the statistical ensemble ${\cal E}[N,{\cal S}]$ of systems ${\cal %
S}$, or even instead of it, one can introduce the statistically averaged
dynamic system $\left\langle {\cal S}\right\rangle $, which is defined
formally as a statistical ensemble ${\cal E}\left[ N,{\cal S}\right] $, ($%
N\rightarrow \infty $), normalized to one system. Mathematically it means
that, if ${\cal A}_{{\rm E}}\left[ N,d_{N}\left\{ X\right\} \right] $ is the
action for ${\cal E}\left[ N,{\cal S}\right] $, then 
\[
\left\langle {\cal S}\right\rangle :\;\;{\cal A}_{\left\langle
S\right\rangle }\left[ d\left\{ X\right\} \right] =\lim_{N\rightarrow \infty
}\frac{1}{N}{\cal A}_{{\rm E}}\left[ N,d_{N}\left\{ X\right\} \right]
,\qquad d\left\{ X\right\} =\lim_{N\rightarrow \infty }d_{N}\left\{
X\right\} 
\]
is the action for $\left\langle {\cal S}\right\rangle $, where $X$ is a
state of a single system ${\cal S}$, and $d_{N}\left\{ X\right\} $ is the
distribution, describing in the limit $N\rightarrow \infty $ both the state
of the statistical ensemble ${\cal E}\left[ N,{\cal S}\right] $ and the
state of the statistically averaged system $\left\langle {\cal S}%
\right\rangle $.

Replacement of the statistical ensemble ${\cal E}\left[ N,{\cal S}\right] $
by the statistically averaged system $\left\langle {\cal S}\right\rangle $
is founded on the insensibility of the statistical ensemble to the number $N$
of its elements, under condition that $N$ is large enough. The statistically
averaged system $\left\langle {\cal S}\right\rangle $ is a kind of a
statistical ensemble. Formally it is displayed in the fact that the state of 
$\left\langle {\cal S}\right\rangle $, as well as the state of the
statistical ensemble ${\cal E}\left[ N,{\cal S}\right] $ is described by the
distribution $d_{N}\left\{ X\right\} $, $N\rightarrow \infty $, whereas the
state of a single system ${\cal S}$ is described by the quantities $X$, but
not by their distribution. Using this formal criterion, one can distinguish
between the individual dynamic system ${\cal S}$ and the statistically
averaged system $\left\langle {\cal S}\right\rangle $.

To obtain the quantum mechanics as a statistical description of stochastic
motion of micro particles, one needs to make one important step more. It is
necessary to introduce the wave function $\psi $, which is the main object
of quantum mechanics. Usually the wave function is introduced axiomatically,
i.e. as an object, satisfying a system of axioms (principles of quantum
mechanics). For this reason the meaning of the wave function is obscure. To
clarify it, one has to introduce the wave function as an attribute of some
model.

If ${\cal S}$ is a particle (deterministic or random), then the statistical
ensemble ${\cal E}\left[ N,{\cal S}\right] $ of particles ${\cal S}$, or
statistically averaged particle $\left\langle {\cal S}\right\rangle $ are
continuous dynamic systems of the fluid type. It is well known \cite{M26},
that the Schr\"{o}dinger equation can be represented as an equation,
describing irrotational flow of some ideal fluid. In other words, the wave
function can be considered to be an attribute of irrotational fluid flow.
One can show \cite{R99}, that the reciprocal statement (any fluid flow can
be described in terms of a wave function) is also valid. The rotational flow
is described by a many-component wave function. In other words, at the
rotational flow the spin appears.

As far as the statistically averaged particle $\left\langle {\cal S}%
\right\rangle $ is a fluid, the wave function appears to be a description
method of this fluid $\left\langle {\cal S}\right\rangle $. For the
statistical description of the particle ${\cal S}$  coincides with the
quantum mechanical description, it is necessary to find the state equation
of the fluid $\left\langle {\cal S}\right\rangle $, which is determined in
turn by the form of the distortion function $D$. Corresponding calculation
was made in the paper \cite{R91}. This calculation determines the form (\ref
{a1.6}) of the distortion function. Then one obtains the conception, which
will be referred to as the model conception of quantum phenomena (MCQP). For
the conventional presentation of quantum mechanics the term ''the axiomatic
conception of quantum phenomena'' (ACQP) will be used.

Dynamical conception of statistical description (DCSD) generates a less
informative description, than the probabilistic statistical description in
the sense that some conclusions and estimations, which can be made at the
probabilistic description, cannot be made in the scope of DCSD. One is
forced to accept this, because one cannot obtain a more informative
description. The fact that the quantum mechanics is perceived as a dynamical
(but not as a statistical, i.e. probabilistic) conception is connected with
the employment of DCSD. In turn application of DCSD is conditioned by
''relativistic roots'' of the nonrelativistic quantum mechanics. The
''dynamic perception'' of quantum mechanics takes place in the scope of both
conceptions MCQP and ACQP. Let us note that DCSD is an universal conception
in the sense that it can by used in both relativistic and nonrelativistic
cases.

\section{Identification of individual particle with the \newline
statistically averaged one}

''Dynamical perception'' of quantum mechanics leads to the fact that the
statistically averaged particle $\left\langle {\cal S}\right\rangle $,
described by the wave function, is considered to be simply a real particle $%
{\cal S}$. The question, why the real particle ${\cal S}$ is described by
the wave function $\psi $, i.e. by the continuous set of variables (but not
by position and momentum as an usual particle), is answered usually, that it
is conditioned by the quantum character of the particle. One refers usually
to the quantum mechanics principles, according to which the quantum particle
state is described by the wave function $\psi $, whereas the classical one
is described by a position and a momentum. At this point we meet AD.6, when
one does not differ between the statistically averaged particle $%
\left\langle {\cal S}\right\rangle $ and the individual particle ${\cal S}$.

As a corollary of such an identification the properties of $\left\langle 
{\cal S}\right\rangle $ and ${\cal S}$ are confused, and an object with
inconsistent properties appears. As long as we work with mathematical
technique of quantum mechanics, dealing only with $\left\langle {\cal S}%
\right\rangle $, no contradictions and no paradoxes appear. But as soon as
the measurement process is described, where both objects $\left\langle {\cal %
S}\right\rangle $ and ${\cal S}$ appear, the ground for inconsistencies and
paradoxes come into existence. Combinations of contradictory properties may
be very exotic.

There are at least two different measurement processes. The measurement ($S$%
-measurement), produced under an individual system ${\cal S}$, leads always
to a definite result and does not influence the wave function, which is an
attribute of the statistically averaged system $\left\langle {\cal S}%
\right\rangle $. The measurement ($M$-measurement), produced under the
statistically averaged system $\left\langle {\cal S}\right\rangle $, is a
set of many $S$-measurements, produced under individual systems ${\cal S}$,
constituting the statistically averaged system $\left\langle {\cal S}%
\right\rangle $. The $N$-measurement changes the wave function of the system 
$\left\langle {\cal S}\right\rangle $ and does not lead to a definite
result. It leads to a distribution of results.

The following situation takes place the most frequently. One considers that
the wave function describes the state of an individual system, and a
measurement, produced under individual system, changes the state (wave
function) at this system. As a result a paradox, connected with the wave
function reduction and known as the Schr\"{o}dinger cat, appears. A
corollary of such an approach is so called many-world interpretation of
quantum mechanics \cite{E57,DG73}.

\section{Identification of Hamiltonian and energy at the secondary
quantization of relativistic field}

The energy of a closed dynamic system is defined as the integral from the
time component $T^{00}$ of the energy-momentum tensor 
\begin{equation}
E=\int T^{00}d{\bf x}  \label{a5.1}
\end{equation}
The energy is a very important conservative quantity. The Hamilton function
(Hamiltonian) of the system is a quantity canonically conjugate to the time,
i.e. the quantity, determining the time evolution of the system. By their
definitions the Hamiltonian $H$ and the energy $E$ are quite different
quantities. But in the nonrelativistic physics (classical and quantum) these
quantities coincide in many cases. For instance, the energy of a particle in
a given potential field $U\left( {\bf x}\right) $ has the form $E={\bf p}%
^{2}/2m+U\left( {\bf x}\right) $. The Hamiltonian of the particle has the
same form. On the ground of this coincidence an illusion appears, that the
energy $E$ of dynamical system plays a role of the quantity, determining its
evolution, i.e. the role of its Hamiltonian $H$. An illusion appears that
the energy and the Hamiltonian are synonyms, i.e. two different names of the
same quantity.

This identification of energy and Hamiltonian is used in the relativistic
quantum theory, where such an identification cannot be used. For instance,
it is common practice to consider \cite{S61}, that in the dynamic system $%
{\cal S}_{{\rm KG}}$, described by the Klein-Gordon equation, the particle
energy may be both positive and negative. A ground for such an statement is
the fact that the flat wave in ${\cal S}_{{\rm KG}}$ has the form 
\begin{equation}
\psi =Ae^{ik_{0}t-i{\bf kx}}  \label{a5.2}
\end{equation}
where the quantity $k_{0}=\sqrt{m^{2}+{\bf k}^{2}}$ is interpreted as an
energy. It may be both positive and negative. The statement that the energy
may be negative is made in spite of the fact that the energy-momentum tensor
component 
\begin{equation}
T^{00}=m^{2}\psi ^{\ast }\psi +{\bf \nabla }\psi ^{\ast }{\bf \nabla }\psi
\label{a5.3}
\end{equation}
which enters in the expression (\ref{a5.1}), takes only nonnegative values.
In reality, the quantity $k_{0}$ is a time component of the canonical
momentum, which can have any sign. But the particle energy is always
nonnegative.

Thus, in the given case one has the associative delusion (AD.7), which lies
in the fact that the properties of Hamiltonian are attributed to the energy.
As long as such an identification is produced on the verbal level, it leads
only to a confusion in interpretation and nothing more. But in the quantum
field theory (QFT) such an identification has a mathematical form, and it
has far-reaching consequences for the secondary quantization of the scalar
field $\psi $. In the relation 
\begin{equation}
i\hbar \frac{\partial \psi }{\partial t}=H\psi -\psi H,  \label{a5.7}
\end{equation}
describing evolution of the dynamic variable $\psi $ in the Heisenberg
representation, the Hamiltonian $H$ is replaced by the energy (\ref{a5.1}),
i.e. the relation (\ref{a5.7}) is written in the form 
\begin{equation}
i\hbar \frac{\partial \psi }{\partial t}=E\psi -\psi E  \label{a5.6}
\end{equation}
The relation (\ref{a5.6}) is an additional constraint which is not necessary
for carrying out the secondary quantization. (The secondary quantization may
be produced without imposing the condition (\ref{a5.6})). If the additional
condition is imposed, one should test its compatibility with dynamic
equations. Unfortunately, there is no understanding of such a test
necessity. The condition (\ref{a5.6}) appears to be compatible with the
dynamic equations only in the trivial case of the linear field $\psi $. For
the nonlinear field the condition (\ref{a5.6}) appears to be incompatible
with dynamic equations that manifests itself as the form of nonstationary
vacuum state.

Theory of the scalar field $\psi$, quantized in accord with the condition (%
\ref{a5.6}), is an inconsistent conception, which is convenient in the
relation that it (as any inconsistent conception) admits one to explain some
facts, which cannot be explained in the scope of a consistent conception,
which does not use the constraint (\ref{a5.6}).

The identification of the energy and Hamiltonian $E=H$ (condition (\ref{a5.6}%
)) has been used in QFT during the second half of XX century, i.e. about 50
years. This condition is not necessary for the secondary quantization, and
its incompatibility with dynamic equations is rather evident. Why was this
condition not come under a storm of criticism in this long time? In QFT
there are many difficulties and problems, and the QFT consistency was open
to question many times. But the condition (\ref{a5.6}) had not to be in
doubt. The author had discussed this question with experts in QFT, but any
time the their reaction was similar. The author opponents did not adduce
counter-arguments and agreed that the condition (\ref{a5.6}) is not too
well, but they rejected to make conclusions, concerning the second
quantization procedure. Put very simply, they searched for reasons of the
QFT difficulties on its surface, giving up to analyze the QFT foundations.

\section{On styles of investigation}

Their considerations look approximately as follows. Let us introduce an
additional supposition and study its consequences for theory and experiment.
If the consequences are positive, the additional supposition is accepted and
introduced into the theory. If the consequences are negative, the additional
supposition is removed and a new additional supposition is considered. Such
additional suppositions were: normal ordering, renormalizations, increase of
the space-time dimension with the subsequent compactification, strings, etc.
This style of investigation: additional supposition with subsequent test of
its consequences will be referred to as P-style (pragmatic style) of
investigation. Such a style is characteristic not only for the QFT
development. In the beginning of XX century the quantum mechanics
development was carried out also by means of P-style. The quantum mechanics
developed, fighting against the classical style (C-style) of investigations,
established to the end of XIX century. In this fight the P-style gained a
victory over the C-style, which played a role of representative of classical
(nonquantum) physics. Successors of Ptolemeus used the P-style, whereas
successors of Copernicus used the C-style. The competition of successors of
Ptolemeus with the successors of Copernicus was at the same time a
competition between P-style and C-style. Then the C-style gained the
victory. C-style reached its fullest flower to the end of XIX century. At
the investigations of quantum phenomena in the XX century C-style gave the
way to P-style.

Why do two different styles of investigation exist? Why does the
investigation C-style or the investigation P-style gain alternatively the
competition? The answer is as follows.

C-style is a style of investigations in the scope of a consistent theory. It
puts in the forefront the consistency of a theory. C-style restricts
suggestion of additional suppositions (hypotheses), insisting, that
additional suppositions be consistent with primary principles of a theory.
(Let us recall the Newton's words: ''I do not invent hypotheses''). In
virtue of its requirement rigidity the C-style has the more predictable
force, than the P-style, where these requirements are not so rigid. Among
the C-style requirements there are ethic requirements to researchers. For
instance, an researcher, which publishes insufficiently founded paper,
containing arbitrary (i.e. not following from the primary principles)
suppositions, risks losing his scientific face.

Adherents of the C-style pay attention to fundamental problems of a theory,
and in particular, to results and predictions of the theory, which are
important for its further development. Solutions of concrete practical
problems are considered to be not so important, because a solution of any
special problem is a formal application of primary principles and
mathematical technique to conditions of the new problem, and nothing beyond
this. Such a relation of the researcher, using the C-style, to a solution of
special problems is founded on his confidence that the primary principles
are valid and the theory is consistent.

The predictability of the C-style, rigidity of its requirements and its
self-reliance are true, provided the primary principles of a theory are
true. If the primary principles contain a mistake, some predictions of the
theory appears to be false. It forces onto searching for a mistake, which
may occurs in the primary principles or in the conclusion of corollaries
from them. The most frequently a mistake is discovered in incorrect
application of the primary principles.

But if the mistake in conclusions of a theory (discrepancy between
predictions of the theory and experiment) has not been discovered for a long
time, the necessity of the cognition progress and necessity of improvement
of the terminology for the experimental data description generate a more
pragmatic style (P-style) of investigations.

The P-style puts in the forefront a possibility of the experimental data
explanation, what is obtained usually by introduction of additional
suppositions. The theory consistency is considered to be not so important.
although the representatives of the P-style declare, that they tend to
elimination of inconsistencies, but it does not succeeded always, and is
considered to be a less defect, than impossibility of the experiment
explanation. The P-style admits an introduction of additional suppositions,
even if they appear to be inconsistent with primary principles. It is
important only, that they were useful and led to explanation of experimental
data. The P-style imposes essentially more slight requirements to
researchers. For instance, the scientific reputation of a researcher does
not lack or lacks slightly, if he, writing a very good paper, writes
thereafter several mediocre or even incorrect papers. Predictability of the
P-style is essentially less, than that of the C-style, as far as P-style
admits only a ''short logic'' (short logical chain of considerations). For
instance, it is widely believed among researchers dealing with quantum
theory that essentially new result can be obtained, only suggesting some
essentially new supposition in the scope of quantum theory. The idea that a
novelty may be found in the primary principles (i.e. outside the scope of
quantum theory) and the new result is a corollary of a long logical chain of
considerations is perceived as something unreal.

Pragmatism of the P-style manifests itself in setting in the forefront a
solution of concrete practical problems. It is supposed that a young talent
gifted researcher is to solve concrete problems, whereas solution of
fundamental problems is supposed to be a work for elderly experienced
researchers. According to such a viewpoint usually one ignores and does not
discuss facts and results which are important for further development of a
theory, but which do not deal directly with its practical applications.
Behind such a relation one can see an uncertainty of the P-style
representatives in the primary principles of a theory and in its
consistency. If a practical problem fails to be solved, the P-style
representatives are ready to suggest additional suppositions and even to
revise the primary principles.

The P-style appears to be more effective, only if the C-style appears to be
ineffective. The last takes place, if the primary principles contain either
mistake or defect. In other words, the C-style is more effective, than the
P-style only at absence of obstacles (systematic noise). The P-style is
noise-resistant, under presence of the ''systematical noise'' it appears to
be more effective, than the C-style. In the period of a long P-style
dominance a theory degenerates. Accumulating many additional supposition,
contradicting each other, the theory gives up step-by-step its predictable
force and capacity of valid development. Situation was such in the time of
dominance of the Ptolemaic doctrine. The same situation takes place now in
the quantum field theory.

In general, the C-style is more effective and predictable, provided the
primary principles are valid. The P-style is useful in the relation, that it
works even in the case, when there is a mistake in the primary principles,
and C-style cannot work. In this case the P-style admits one to introduce
new adequate concepts and terminology for descriptions of experiments that
cannot be explained by the theory, based on the primary principles. Finally,
investigations, realized by means of the P-style, help one to discover
mistake in the choice of primary principles and produce a necessary revision.

Any style of investigations is conservative. It is worked out by a
researcher in the course of all his research activity. If the researcher
used the P-style, i.e. he uses essentially the trial and error method, he
gets accustomed hardly to rigid restrictions of the C-style. Vice versa, a
researcher, using the C-style in his work, gets accustomed to work with
consistent conceptions. It is very difficult for him to pass to more free
P-style and to invent new additional supposition which are necessary for
explanations of new experiments. Conservatism of the investigation style
leads to a conflict, when the dominating investigation style changes. For
instance, in the time of Ptolemeus the P-style dominated. Discovery of AD.2
needed to construct a consistent conception of the celestial mechanics which
would be free of arbitrary suppositions. The conflict between the successors
of Ptolemeus and those of Copernicus was in the same time a conflict between
the investigation styles.

Now practically all researchers dealing with relativistic QFT use P-style.
They perceive difficultly arguments of the C-style proponents, having found
inconsistencies and mistakes in primary principles of the quantum theory.

\section{History of the associative delusions overcoming}

Associative delusions have rather unusual properties. Usually they are
overcame unintentionally. As a rule, the associative delusions are
discovered and realized only after they have been overcame. Overcoming of
AD.2 (conflict between the Copernicus doctrine and the Ptolemaic one)
continues for several decades. Both sides of the conflict understood very
well the reason of disagreement, but the fact, that this reason was an
associative delusion, was interesting for nobody.

The following associative delusion AD.3 (the Cartesian coordinate system as
a fundamental object of geometry) had been overcame in the course of the
second half of XIX century. Its overcoming appeared to be not noticed in the
sense that the reason of conflicts in the scientific community remained to
be obscure. In the case of AD.2 it was clear, what was a subject of
disagreement. In the case of AD.3 the reason of the conflict remained to be
not clear. The mathematical community did not accept non-Euclidean geometry.
There is the evidence of Felix Klein \cite{K37}. It is known also that the
Russian mathematical community related with a prejudice to works of
N.I.~Lobachevski on the geometry \ of negative\  curvature and to
N.I.~Lobachevski himself as an author of these works \cite{M81}.
Unfortunately, the reasons of this prejudice had not been analyzed (at any
case such an analysis is not known for us), and this episode remained in the
history of mathematics as an unclear flash of conservatism.

Associative delusions AD.4 -- AD.7 are found at the stage of overcoming. On
one hand, an objective analysis of methods of AD overcomings is very
difficult. On the other hand, such an analysis is necessary, because it
would make overcoming of AD.4 -- AD.7 easier. Besides AD.4 -- AD.7 are not
the last associative delusions on the way of the cognition process.
Investigation of properties of associative delusions could help one to
overcome the next AD.

AD.4 -- AD.7 were discovered and overcame by one person -- the author of
this paper. Discovery and overcoming of ADs is rather rare processes. In
this connection it is very difficult to answer the questions of the type.
With what is a discovery of ADs connected? What are the circumstances, at
which the discovery of ADs takes place? Why do some researchers succeed in
overcoming of ADs, whereas other do fail?

Motives and attendant circumstances are known usually only for those
researchers, who participated directly in overcoming of AD (in the given
case this is the author of the paper). The process of overcoming was rather
long (about 40 years). The diary was not kept, and one is forced to entrust
to recollections. But the human memory is not enough reliable. The human
beings are apt to forget events and circumstances, especially, if they are
unpleasant for them. A subconscious mythologization of the investigation
process takes place. But recollections of the only direct participant of the
process of the ADs discovery and overcoming are of a certain interest for
further investigations of associative delusions, even if they are subjective
and strongly mythologized. Such recollections are of interest for subsequent
investigators even in the case, if their author had not understood, or had
understood incorrectly, what he had done in reality. The recollections are
unique and are of interest, even if their author had described instead of
important circumstances some unessential details, which he remembers for
some reasons.

Experience of overcoming of AD.4 -- AD.7 evidences that in some case the
overcoming of associative delusion happens to be accidental, or it is an
accessory result of investigations, carried out with other goal. In other
cases ADs appear to be connected with insufficient understanding of the
existing theory (in particular, the relativity theory), and overcoming of AD
is carried out consciously. However, taking part in the process of the AD
overcoming, it is very difficult one to analyze objectively this process.
Apparently, the objective analysis will be possible only, when the process
of the AD overcoming has been over. The author can make his contribution to
this analysis only by one way -- he should tell, how {\it from his viewpoint}
the process of this overcoming passed. Of course, one can produce this
analyze on the ground of published papers. But in this case one cannot take
into account motives of the paper writing and the important circumstance,
that sometimes the obtained results differ from those ones that the author
wanted to obtain. Motives and goals of the author are unessential for
estimation of his contribution, but they are essential for investigation of
the cognition process (analyses of the process of the AD overcoming).

In general, recollection should be printed in the memoirs, but not in the
scientific paper. In the given case there is the excuse that the question is
connected not only with recollections, but with recollections accompanied by
formulae. Besides the question is about evidences (maybe, very subjective),
concerning very rare cognition process, which is the AD overcoming.

Thus, evidences of the participant will be presented. They are presented
from the first person singular, in order to underline subjective character
of recollection and separate them from other part of presentation, which
pretends to objectivity. The evidences of participant do pretend by no means
to a review of investigations, connected with overcoming of AD.4 -- AD.7,
and all references to other researchers are cited so far as they influenced
the participant investigations and remained in his memory.

\section{Beginning of AD.4 overcoming. Evidence of participant.}

The idea that the space-time is described completely by interval (distance)
between pairs of events appeared at once after my acquaintance with the
relativity theory. (It took place in 1955, when I was a second-year
student). I supposed that all experts thought the same and did not see
nothing new and surprising in such a point of view. (Some physicists told me
many years ago, that they hold the same viewpoint). I was somewhat surprised
that the infinitesimal interval was used (but not finite). I explained this
fact to myself that it was easier to work with infinitesimal interval,
because it contained less information (functions of one space-time point),
than the finite interval (function of two space-time points). At first, I
did not understand the circumstance, that an introduction of an
infinitesimal interval is impossible without introduction of the dimension
and manifold. I considered the dimension and the manifold to be natural
attributes of the space-time.

Some years later I undertook a study of that, to what extent the space-time
could be described in terms of a finite interval. The finite interval was
attractive by the fact, that it contains more information, than the
infinitesimal one. For instance, a geodesic, described usually by a system
of four ordinary second order differential equations, is described in terms
of the finite interval $S\left( x,x^{\prime }\right) $ as a solution $%
x^{\prime }=x^{i}\left( \tau \right) ,\;\;i=0,1,2,3$ of a system of four
algebraic equations 
\begin{equation}
\frac{\partial G\left( x,x^{\prime }\right) }{\partial x^{\prime i^{\prime }}%
}=b_{i^{\prime }}\tau ,\qquad b_{i^{\prime }}=\text{const,\qquad }G\left(
x,x^{\prime }\right) =\frac{1}{2}S^{2}\left( x,x^{\prime }\right)
\label{a8.1}
\end{equation}
The quantity, which is denoted now via $\sigma $ and known as the world
function, I denoted by $G$. For me this designation associated with
gravitation. The quantities, depending on two space-time points, I denoted
by capital letters.

It should note that I had worked out somewhat unusual and, as I am
understanding now, unpleasant for my colleagues style of work. The style of
my work was such a kind, that before undertaking an investigation I had not
studied or had studied very slightly works of my predecessors. I connect
appearance of such a style with the fact, that in the course of my studying
at the university I had practically no scientific advisor, who could give me
a definite theme for investigations and press on so that I investigated it.
Of course, I had an official scientific advisor. He suggested me a theme for
a diploma work and for the Ph.D. thesis. But I was a self-willed student. I
ignored recommendations of my scientific advisor and investigated those
subjects, which I considered to be interesting and necessary. My scientific
advisor, qualifying me as a ''non-controlled student'' did not insist on his
choice of the subject of investigation and does not hinder my self-will in
the choice of the investigation subject. Moreover, he supported me by all
means in my work.

Usually the scientific advisor suggests the subject for investigation and
related literature. Doing so he guarantees that this work will not be a
rediscovery of known results. I began my investigation of the finite
interval properties (world function), as if such investigations had not been
produced earlier. I knew nothing about them, and nobody of my colleagues
could tell me anything about such investigations. Investigation was made in
spring and in summer of 1958, when I was five-year student of the physical
faculty of the Moscow Lomonosov university.

Understanding possibility of rediscovery of known results, I tended
subconsciously to prevent such a rediscovery. To eliminate the rediscovery,
I must bring my investigation to obtaining of certainly new and interesting
result. The possibility that my intermediate results appeared to overlap
already known results, has not agitated me\footnote{%
I guessed that it is not very ethically not to refer to my predecessors and
tried not to admit this. But the fact that absence of references to
predecessors disoriented the readers, which are acquainted at first with
this subject, did not come to my mind. I shall know about this a bit later.}%
. I ignored the evident way of investigation of the world function $G\left(
x,x^{\prime }\right) $ -- expansion into a series over powers of $%
x-x^{\prime }$ and paid attention on the circumstance that the quantity $%
\Gamma _{kl}^{i}\left( x,x^{\prime }\right) \equiv G^{is^{\prime
}}G_{kl,s^{\prime }}$ is a scalar at the point $x^{\prime }$ and the
Christoffel symbol at the point $x$. Here the following designations are
used
\begin{equation}
G_{kl^{\prime }}\equiv \frac{\partial ^{2}G}{\partial x^{k}\partial
x^{\prime l^{\prime }}},\qquad G_{kl,s^{\prime }}\equiv \frac{\partial ^{3}G%
}{\partial x^{k}\partial x^{l}\partial x^{\prime s^{\prime }}}  \label{a8.2}
\end{equation}
and $G^{is^{\prime }}$ is the matrix reciprocal to the matrix $G_{kl^{\prime
}}$.

Moreover, it appeared that $\Gamma _{kl}^{i}\left( x,x^{\prime }\right) $ is
the Christoffel symbol for a flat space, i.e. the Riemann -- Christoffel
curvature tensor for $\Gamma _{kl}^{i}\left( x,x^{\prime }\right) $ is equal
to zero identically. As a result it appeared that one could introduce such a
covariant derivative with respect to $x^{i}$, which was a covariant
derivative in some flat space and which depended on a parameter -- the point
$x^{\prime }$.

All this meant, that the world function and the Riemannian space $V$, which
was described by it, were associated with a set of flat spaces $E_{x^{\prime
}}$, labelled by the point $x^{\prime }$ of the Riemannian space $V$.
Investigation of this question showed that the spaces $E_{x^{\prime }}$ were
flat spaces tangent to the Riemannian space $V$ at the point $x^{\prime }$.
It appeared that the world function generated automatically some two-metric
technique, realizing a geodesic mapping of the Riemannian space $V$ onto $%
E_{x^{\prime }}$. Moreover, it was discovered later on \cite{R92}, that such
a mapping was realized by any symmetric scalar function $G\left( x,x^{\prime
}\right) $ of two points, for which one could determine the tensor $%
G^{is^{\prime }}$. But in that time I had not known this result and did not
think about a possibility of coming outside the scope of Riemannian geometry.

By the way, results of the work admitted one to make this already then. The
most evident result of the work was a construction of the two-metric
technique on the basis of the world function. It was the result that
impressed my colleagues mostly. I myself considered that the main result of
the paper was an obtaining of a system of differential equations for the
world function of Riemannian space. The equation, which is satisfied by the
world function $\sigma $ of the Riemannian space is well known \cite{S60}.
\begin{equation}
\frac{\partial \sigma }{\partial x^{i}}g^{ik}\left( x\right) \frac{\partial
\sigma }{\partial x^{k}}=2\sigma   \label{a8.3}
\end{equation}
It contains the metric tensor of the Riemannian space in an explicit form,
and it cannot serve as an equation, splitting all symmetric functions $%
G\left( x,x^{\prime }\right) $ into two sets: one set contains the
functions, which can be a world function of a Riemannian space, another set
contains functions $G\left( x,x^{\prime }\right) $, which cannot play this
role.

Using the two-metric technique, one succeeded to eliminate the metric tensor
from the equation (\ref{a8.3}) and obtain a system of differential equations
for the world function, which did not contain metric tensor explicitly, but
in return it contained derivatives of the world function with respect to
both arguments $x$ and $x^{\prime }$ \cite{R62}. The world function of any
Riemannian space satisfied this system of equations.

The system of equations put the question. What is described by the world
function, which does not satisfy this system? Does it describe
non-Riemannian geometry, or no geometry at all? In that time (in the
beginning of sixtieths) I did not put this question, but I considered
derivation of this system of equations as a main result of my work. My
attempts to call attention of my colleagues to this result did not lead to a
success. Agreeing that the development of the two-metric technique was a
progress, they were indifferent to the system of equations for the world
function of a Riemannian space, that was considered by me as the main result.

The paper was published in slightly known journal \cite{R62} in 1962, almost
three years after it had been submitted. In the time between the submission
and publication some very important for me event happened. The book by
J.L.~Synge \cite{S60}, where at first the general relativity was presented
in terms of the world function, appeared in Moscow. I had learned from it
that the world function was introduced at first by H.S.~Ruse \cite{R31} and
J.L~Synge \cite{S31} practically simultaneously, and that its properties had
been investigated.

But the book by J.L.~Synge did not contain any information on two-metric
technique and geodesic mapping on tangent spaces $E_{x^{\prime }}$. My
misgivings that results of my investigations would appear to be a
rediscovery of known results were not justified. Moreover, my study of the
book by J.L.~Synge led to increase of my self-reliance. If being a student,
I could without a help make an investigation and obtained the results, which
had not been obtained by so experienced researcher as J.L.~Synge, it meant
that I was able to compete with any researchers. Further this increase of my
self-reliance will help me in my investigations. This case validated my
style of working, when I did not troubled myself with study of the
literature, postponing this to the time, when the first results deserving
publication would be obtained. I used such a style of work, because I
preferred to develop a new direction, but not to continue investigations of
my predecessors. At first, it was easier, and, second, an attentive study of
literature influenced me very strongly. It pushed me to a beaten way and
prevented from a choice of directions of investigations, natural from my
viewpoint. These directions appeared sometimes to be alternative to
conventional directions of investigations.

Until 1964 I had written several papers about the two-metric technique
applications to gravitation and several papers on the unified field theory.
Dealing with the unified field theory, I was led to conclusion that all
universal phenomena and universal constants were a manifestation of the
space-time properties. For instance, I was sure (and I am sure now) that
multiplicity of any electric charge to elementary one was conditioned by the
space-time properties, as well as the quantum constant $\hbar $ was the
space-time attribute. Besides to the end of 1964 I was convinced that one
cannot succeed in the study of the space-time properties without
understanding what was the quantization. This (and other circumstances)
forced me to leave the Moscow Lomonosov university and to change a subject
of my investigations.

\section{Relativistic statistics and overcoming of AD.5, AD.6. Evidence of
participant.}

Since 1964 I stopped to deal with geometry, gravitation, and unified field
theory. In several years I was dealing with applied physics, and in the
beginning of seventieths I started to develop the direction, which I called
''quantum mechanics as a relativistic statistics''. It is this direction,
whose development is connected with overcoming of AD.5. Then I assumed that
the quantum mechanics was a result of statistical description of randomly
moving particles. It seemed to me very reasonable in the light of the fact
that in the end of XIX century the thermal phenomena were explained
successfully as a result of statistical description of chaotically moving
molecules. I assumed that the quantum phenomena could be explained in the
same way. This idea was not new, but numerous attempts to realize this
program did not led to a success. As far as know, most of researchers
considered this program to be wrong. They assumed that the quantum
properties reflected ''quantum nature'' of the world and that they could not
be reduced to other ''classical'' properties.

I did assume that failure of the statistical description was connected with
the nonrelativistic character of the statistical description. I assumed that
although the quantum mechanics was nonrelativistic, but nevertheless, the
statistical description was to be relativistic. In my opinion, the fact that
the regular component of the particle motion was relativistic did not mean
that the stochastic component was nonrelativistic also. To be on the safe
side, one was to use a relativistic statistical description, which should be
valid independently of the case, whether the stochastic component of motion
be relativistic or not.

Being a student, I studied the relativity theory, reading the book by
V.A.~Fock \cite{F55}, and shared his viewpoint that the relativistic theory
differed from the nonrelativistic one not only in relativistic invariance of
their dynamic equations. As it is well known, the main difference between
the special relativity theory (SRT) and the nonrelativistic physics lies in
the fact that in the nonrelativistic physics there are the concept of
simultaneity in remote points, whereas in SRT there is no such a
simultaneity. It means essentially that the nonrelativistic physics is
incompatible with the relativistic physics. But a compromise between the
relativity and nonrelativistic physics was strongly necessary, as far as the
physics remained to be mainly nonrelativistic. Presentation of the whole
nonrelativistic physics in the relativistic form, to take only into account
small relativistic corrections, seemed to be unacceptable. One found a
compromise. A concept of the relative simultaneity (simultaneity on a given
coordinate system) was introduced. Thus, on one hand, the simultaneity was
considered as though it were, and nonrelativists were satisfied. On the
other hand the simultaneity was relative. It was considered as though it was
not exist, and relativists were satisfied also. On one hand, for
nonrelativists this compromise facilitated a transition to relativistic
physics, admitting them to conserve nonrelativistic style of thinking at
description of relativistic processes. But on the other hand, it was this
style of thinking that hindered investigation of physical phenomena, whose
relativistic character was not evident.

Unfortunately, this restricted character of the compromise, connected with
application of the relative simultaneity, was understood only by a few
physicists. Practically all textbooks on the relativity theory were written
on the ground of this compromise. Apparently, it is this compromise that is
a reason why most of researches ignored application of space-time diagram in
their investigations. Usually description was produced in terms of
coordinate systems and relativistic invariance of dynamic equations. In
other words, most of researchers thought in terms of Newtonian mechanics
with corrections for relativistic invariance. Such an approach did not help
to investigate relativistic phenomena, and J.L.~Synge wrote about this in
the preface to his book \cite{S60}. As for me, as well as J.L.~Synge I
thought in terms of space-time diagrams, i.e. in terms of space-time
geometry.

I assumed that the probability theory could not be used for statistical
description of relativistic stochastic particles, as far as concept of the
particle state is different in SRT and in the nonrelativistic physics. In
SRT the physical object, connected with the particle is WL (world line). The
fundamental concept, connected with the particle, is WL. The WL is primary,
and a particle is an attribute of WL. The particle is a section of WL by the
hyperplane $t=$const. These sections are different in different coordinate
systems, especially if there are several WLs.
In the nonrelativistic physics a particle is a physical object, and world
line is its attribute (history). 

The most difference between the two approches appears at description of
the pair production and pair annihilation. If the physical object is WL,
the turn of world lines in the time direction, when the time coordinate 
$x^0$ is not a monotone function of the parameter $\tau$ along WL, does 
not provoke objections. But if the physical object is a particle, which is 
described in the terms of some dynamic system, an annihilation of the 
particle together with the dynamic system describing it is considered 
to be an absurd from the viewpoint of classical mechnics principles. The 
classical mechanics does not consider production and annihilation of 
dynamic systems, and the pair production process can be understood only 
from viewpoint of quantum mechanics.

Description of a physical object is divided
into two parts: (1)~the object state and (2)~dynamic equations, describing
the state evolution. Dynamics is insensitive to the method of decomposition
into state and dynamic equations. For dynamics it is of no importance what
is a physical object: WL, or a particle. Vice versa, the statistical
description is a calculus of states, and it is very important for it, what
is the physical object and what is a state of this object. Different choice
of the physical object and different decomposition into state and dynamic
equations lead to different results.

The probability theory is suitable for calculus of the particle states, but
not for states of WLs. But, it is a long way from understanding, that the
probability theory is incompatible with the statistical description of
relativistic particles, up to a construction of statistical description
without probability.

Understanding that the density of states of world lines is described by the
4-vector $j^{i}$, I tried to formulate the quantum mechanics in terms of
4-vector $j^{i}$, i.e. to obtain that which was called hydrodynamical
interpretation of quantum mechanics. I tried to develop a mathematical
technique, which could convert the state density $j^{i}$ into a fundamental
object of a theory, making the wave function to be an attribute of this
object. I succeeded to state the statistical principle, determining
construction of the statistical description \cite{R80}. Essentially, the
statistical description was formulated as follows. At the statistical
description one considers many independent stochastic particles. They form
something like a gas or a fluid, whose state is described by the mean 4-flux
of particles $j^{i},\quad i=0,1,2,3$. Thus, one overcame AD.5 (consisting in
the statement that a use of the probability theory is a necessary condition
of the statistical description). To derive the quantum mechanics it was
necessary to pass from $j^{i}$ to the wave function $\psi $. At this point a
purely mathematical problem arose.

The fact is that the transition from the wave function to a hydrodynamics
with the vector $j^{i}$, is well known \cite{M26}. But it leads to a
irrotational flow, i.e. to a very special case of the flow. It was known how
to construct the wave function for the irrotational flow. But how to make
this in the case of arbitrary flow? It was a complicated mathematical
problem, which had taken a long time and many efforts. But finally it had
been solved. It appeared that it was necessary to integrate the complete
system of hydrodynamic equations, i.e. the system, consisting of the Euler
equations and dynamic equations, describing displacement of the fluid
particles in the given velocity field (so called Lin constraints). As a
result of this integration the order of the system reduced, arbitrary
integration functions appeared, and the system became to be described in
terms of Clebsch potentials. Thereafter one could introduce many-component
wave function, which appeared as an attribute of the fluid description.

This result was obtained in the middle of eightieths, and was applied at
first to hydrodynamics \cite{R989}. Thereafter the mathematical technique of
dynamical conception of statistical description (DCSD) appeared. The wave
functions turned to an attribute of the statistical description, and many
unsolved problems became solvable.

It should note here, that there were numerous attempts to revise and to
interpret the quantum mechanics in other way (see. \cite
{M26,B26,B73,BH89,H93} and references therein). All authors of these papers
started from the wave function and Schr\"{o}dinger equation, trying to give
another more acceptable meaning to quantum quantities. As a result they
obtained different interpretations of the quantum mechanics and nothing
more, because the wave function remained to be a fundamental object of a
theory. Turning the wave function to an attribute of statistical
description, I deprived it the status of a fundamental object. As a result
another conception, but not simply another interpretation of quantum
mechanics appeared.

The solved problem was a problem of the hydrodynamics (mechanics of
continuous medium) The hydrodynamics was such a field of physics, where I
was not an expert. In that time I was working at the laboratory, where
practically all researchers were hydrodynamicists. I reported my work on
hydrodynamics at the session of a seminar of our laboratory. Hearing my
report, hydrodynamicists of our laboratory were puzzled, because they did
not see a result in my report. There, where I saw integration with
appearance of three arbitrary functions, they saw only a complicated change
of variables. There was a reason for such a viewpoint. I present briefly an
essence of the problem, because it seems to me very important for
understanding, what is the delusion and how it have been overcame.

Motion of ideal compressible fluid is described usually by five Euler
equations for five dependent variables velocity ${\bf v}$, density $\rho $
and entropy $S$
\begin{equation}
\frac{\partial \rho }{\partial t}+{\bf \nabla }(\rho {\bf v})=0  \label{b1.1}
\end{equation}
\begin{equation}
\frac{\partial {\bf v}}{\partial t}+({\bf v}{\bf \nabla }){\bf v}=-\frac{1}{%
\rho }{\bf \nabla }p,\qquad p=\rho ^{2}\frac{\partial E}{\partial \rho }
\label{b1.2}
\end{equation}
\begin{equation}
\frac{\partial S}{\partial t}+({\bf v}{\bf \nabla })S=0  \label{b1.3}
\end{equation}
Here $p$ is the pressure, and $E=E(\rho ,S)$ is the internal energy per unit
mass, considered to be a function of density $\rho $ and entropy $S$. The
internal energy $E=E(\rho ,S)$ is a unique characteristic of the ideal fluid.

On one hand, the Euler equations form a closed system of dynamic equations,
but on the other hand, they do not form a complete system of dynamic
equations, describing dynamic system ${\cal S}_{{\rm fl}}$, called ideal
fluid. To obtain a complete description, one needs to add more three
equations, describing displacement of the fluid particles in the given field
of velocity ${\bf v}(t,{\bf x})$. It can be made, adding so called Lin
constraints \cite{L63}
\begin{equation}
\frac{\partial {\bf \xi }}{\partial t}+({\bf v}{\bf \nabla }){\bf \xi }=0,
\label{b1.4}
\end{equation}
where Lagrangian variables ${\bf \xi }={\bf \xi }(t,{\bf x})=\{\xi _{\alpha
}(t,{\bf x})\}$, $\alpha =1,2,3$ are considered to be functions of
independent variables $t,{\bf x}$. If one solves the equations (\ref{b1.4})
and determine ${\bf \xi }$ as functions of $(t,{\bf x})$, the finite
relations
\begin{equation}
{\bf \xi }(t,{\bf x})={\bf \xi }_{{\rm in}}=\hbox{const }  \label{b1.5}
\end{equation}
describe implicitly the particle trajectories and motion of particles along
them.

The Lagrangian coordinates ${\bf \xi }$ label the fluid particles, as far as
they represent three independent integrals of the system of ordinary
differential equations.
\begin{equation}
\frac{d{\bf x}}{dt}={\bf v}(t,{\bf x}),\qquad {\bf x}=x(t,{\bf \xi }),
\label{b1.6}
\end{equation}
describing the particle displacement in the given velocity field. The Lin
constraints (\ref{b1.4}) are equivalent to the system (\ref{b1.6}) of three
ordinary differential equations, and hydrodynamicists do not consider them
seriously, supposing that the main problem is a solution of the Euler
equations. If they are solved, i.e. if the system of five partial
differential equations has been solved, the subsequent solution of the
system of three ordinary differential equations is essentially much easier
problem. Essentially, there is no necessity to solve equations (\ref{b1.6}),
as far as the system of Euler equations is closed, and one can find the
fluid flow independently of equations (\ref{b1.6}). The displacement of
fluid particles in itself, described by the equations (\ref{b1.6}), is not
interesting for hydrodynamicists. All this led to that situation, when
hydrodynamicists ignore the Lin constraints both in the form (\ref{b1.4})
and in the form (\ref{b1.6}). In action it led to the fact, that one tried
to write the variational principle for the closed system of equations (\ref
{b1.1}) -- (\ref{b1.3}), but not for the dynamic system ${\cal S}_{{\rm fl}}$%
. It is impossible to write the variational principle for the closed system
of equations (\ref{b1.1}) -- (\ref{b1.3}). One succeeded to make this only
after adding the Lin constraints (\ref{b1.4}) \cite{S88}.

The complete system of dynamic equations (\ref{b1.1}) -- (\ref{b1.4}) is
invariant with respect to the relabelling of the fluid particles. It means,
that the relabelling transformation of the fluid particles
\begin{equation}
\xi _{\alpha }\rightarrow \tilde{\xi}_{\alpha }=\tilde{\xi}_{\alpha }({\bf %
\xi }),\qquad D\equiv \det \parallel \partial \tilde{\xi}_{\alpha }/\partial
\xi _{\beta }\parallel \neq 0,\qquad \alpha ,\beta =1,2,3  \label{b1.16}
\end{equation}
does not change the form of the system of equations (\ref{b1.1}) -- (\ref
{b1.4}), i.e. the relabelling group (\ref{b1.16}) is a symmetry group of the
complete system of equations (\ref{b1.1}) -- (\ref{b1.4}). Application of
the curtailed system of equations (\ref{b1.1}) -- (\ref{b1.3}) admits one to
ignore labelling ${\bf \xi }$ of the fluid particles and carry out the
description in terms of relabelling-invariant variables $\rho ,{\bf v},S$.

There is a method of simplifying the complete system of dynamic equations (%
\ref{b1.1}) -- (\ref{b1.4}), other, than ignoring the variables ${\bf \xi }$
and description in terms of relabelling-invariant variables $\rho ,{\bf v},S$%
. Using the fact that the relabelling group (\ref{b1.16}) is a symmetry
group of dynamic equations, one can integrate the complete system of dynamic
equations (\ref{b1.1}) -- (\ref{b1.4}) in the form
\begin{equation}
S(t,{\bf x})=S_{0}({\bf \xi })  \label{b1.9}
\end{equation}
\begin{equation}
\rho (t,x)=\rho _{0}({\bf \xi })\frac{\partial (\xi _{1},\xi _{2},\xi _{3})}{%
\partial (x^{1},x^{2},x^{3})}\equiv \rho _{0}({\bf \xi })\frac{\partial (%
{\bf \xi })}{\partial ({\bf x})}  \label{b1.10}
\end{equation}
\begin{equation}
{\bf v}(t,{\bf x})={\bf u}(\varphi ,{\bf \xi },\eta ,S)\equiv {\bf \nabla }%
\varphi +g^{\alpha }({\bf \xi }){\bf \nabla }\xi _{\alpha }-\eta {\bf \nabla
}S,  \label{b1.11}
\end{equation}
where $S_{0}({\bf \xi })$, $\rho _{0}({\bf \xi })$ and $g({\bf \xi }%
)=\{g^{\alpha }({\bf \xi })\}$, $\alpha =1,2,3$ are arbitrary functions of
the argument ${\bf \xi }$, and $\varphi $, $\eta $ are new dependent
variables, satisfying the dynamic equations
\begin{equation}
\frac{\partial \varphi }{\partial t}+{\bf u}(\varphi ,{\bf \xi },\eta ,S)%
{\bf \nabla }\varphi -\frac{1}{2}[{\bf u}(\varphi ,{\bf \xi },\eta ,S)]^{2}+%
\frac{\partial (\rho E)}{\partial \rho }=0  \label{b1.12}
\end{equation}
\begin{equation}
\frac{\partial \eta }{\partial t}+{\bf u}(\varphi ,\xi ,\eta ,S){\bf \nabla }%
\eta =-\frac{\partial E}{\partial S}.  \label{b1.13}
\end{equation}
If five dependent variables $\varphi $, ${\bf \xi }$, $\eta $ satisfy the
system of equations (\ref{b1.4}), (\ref{b1.12}), (\ref{b1.13}), the five
dynamic variables $S$, $\rho $, ${\bf v}$ (\ref{b1.9})--(\ref{b1.11})
satisfy the dynamic equations (\ref{b1.1})--(\ref{b1.3}). The indefinite
functions $S_{0}({\bf \xi })$, $\rho _{0}({\bf \xi })$, and $g^{\alpha }(%
{\bf \xi })$ can be determined from initial and boundary conditions in such
a way, that the initial and boundary conditions for variables $\varphi $, $%
{\bf \xi }$, $\eta $ would be universal in the sense that they do not depend
on the fluid flow.

The last condition means that at the description in terms of hydrodynamic
potentials the total information on the fluid flow is contained in dynamic
equations, and there is no necessity in giving the initial and boundary
conditions, because they may be universal. At the description in terms of
relabelling-invariant variables $S$, $\rho $, ${\bf v}$ one needs to add
initial and boundary conditions to dynamic equations to obtain all
information on the fluid flow. The circumstance that dynamic equations (\ref
{b1.12}), (\ref{b1.13}) contain initial and boundary conditions, introduced
by means of (\ref{b1.11}), is unique. In accord with (\ref{b1.10}), (\ref
{b1.11}) the physical quantities $\rho $, ${\bf v}$ are obtained as a result
of differentiation of variables $\varphi $, ${\bf \xi }$, $S$, and the
variables $\varphi $, ${\bf \xi }$, $\eta $ may be interpreted as
hydrodynamic potentials. This way of description can be called as
description in terms of hydrodynamic potentials. These potentials associate
with the name of Clebsch \cite{C57,C59}, who has introduced them for a
description of incompressible fluid.

Thus, ignoring the potentials ${\bf \xi }$, one came to a description in
terms of relabelling-invariant variables $\rho ,{\bf v},S$ by means of five
equations (\ref{b1.1})--(\ref{b1.3}). Description in terms of hydrodynamic
potentials $\varphi ,{\bf \xi },\eta $ led to a description by means of five
equations (\ref{b1.4}), (\ref{b1.12}), (\ref{b1.13}). It seems that the
system of equations (\ref{b1.1})--(\ref{b1.3}) on one hand, and equations (%
\ref{b1.4}), (\ref{b1.12}), (\ref{b1.13}) on the other hand, can be derived
one from other by means of a change of variables. The circumstance, that the
change of variables was differential in one direction, and it was integral
in other direction, was not taken into account.

However, it is very important circumstance. Differentiation of potentials $%
\varphi ,{\bf \xi },\eta $ at the change of variables leads to a loss of
information on the fluid particle displacements, described by the equations (%
\ref{b1.4}). The wave function contains complete information on motion and
displacement of the fluid particles, and it can be built only of potentials $%
\varphi ,{\bf \xi },\eta $, and it cannot be built of relabelling-invariant
variables $\rho ,{\bf v},S$. Understanding of this circumstance appeared to
be difficult for me and for other researchers. The last manifested itself in
the fact that, when (in 1998) my paper on integration of hydrodynamic
equations was submitted to Journal of Fluid Dynamics. It was declined on the
ground that in the reviewer's opinion the transition from the curtailed
system of dynamic equations (\ref{b1.1})--(\ref{b1.3}) to the complete
system (\ref{b1.4}), (\ref{b1.12}), (\ref{b1.13}) does not contain an
integration, but only a change of variables.

Why did I succeeded to construct a fluid description in terms of the wave
function and to solve the problem of hydrodynamics, where I was an undoubted
amateur? I reflected on this question and came to the following conclusion.
At first, I knew what I do search for and had made many unsuccessful
attempts. In the course of these attempts I developed and improved the
mathematical technique of work with Jacobians (Jacobian technique). Looking
some years ago through papers of Clebsch \cite{C57,C59}, I have discovered
that he also used the Jacobian technique and had discovered hydrodynamic
potentials. He wrote down Jacobians in the expanded form, as it was usual in
that time. Due to this the formalism was rather bulky, but I think, that
Clebsch was able to discover description in terms of wave functions, if
there were a necessity for this. Second, I investigated the fluid simply as
a dynamic system, as far as I did not know hydrodynamics. In return, I was
not burdened with prejudices, connected with a knowledge of hydrodynamics.
In particular, in hydrodynamics the variables ${\bf \xi }$ are well known as
Lagrangian coordinates, which are used only as independent variables. I had
met the Lagrangian coordinates as dependent variables only once in the book
by V.A.~Fock \cite{F55}, when I studied the relativity theory, being a
second-year student. I remember very well, that I was surprised by
capacities, included in such a description.

Returning to the problem of overcoming of AD.5 and AD.6, I should like to
note, that both delusions were overcame practically in the beginning of
seventieths. Having perceived, that the statistical description can and must
be constructed on a dynamical basis (without a use of the probability
theory) meant overcoming of AD.5. Overcoming of AD.6 was a corollary of the
fact, that the quantum mechanics is a statistical conception (statistical
description of randomly moving particles).

\section{Forced identification $E=H$. Evidence of \newline
participant on overcoming of AD.7}

Overcoming of AD.7 (Forced identification $E=H$) was the first of my
overcomings of AD. It took place in 1970. It was carried out consciously on
basis of understanding that in the relativistic theory a physical object is
world line (WL)\footnote{%
designations WL is used for the world line, considered as a fundamental
object}, but not a pointlike particle in the three-dimensional space. I took
this truth from the book of V.~A.~Fock. \cite{F55}. Later I found
confirmation of this viewpoint in papers of Stueckelberg \cite{S42} and
Feynman \cite{F49}. In general such a viewpoint was in keeping with my style
of geometrical thinking. This brought up the question: ''Is it possible to
describe pair production in terms of classical relativistic mechanics?'' The
pair production process is described by a turn of a world line in the time
direction. It was well known. It was necessary to invent such an external
field which could carry out this turn. It was clear, that adding an
arbitrary field to the action of charged particle in a given electromagnetic
field $A_{i}$ 
\begin{equation}
{\cal A}\left[ q\right] =\int \left\{ -mc\sqrt{g_{ik}\dot{q}^{i}\dot{q}^{k}}+%
\frac{e}{c}A_{i}\dot{q}^{i}\right\} d\tau ,\qquad \dot{q}\equiv \frac{d\dot{q%
}}{d\tau }  \label{a10.1}
\end{equation}
one could not carry out such a turn. The fact is that, at the turn in time
the world line becomes to be spacelike near the turning point. On the other
hand, under the sign of radical in (\ref{a10.1}) must be a nonnegative
quantity. It means, that $g_{ik}\dot{q}^{i}\dot{q}^{k}\geq 0$ and, hence the
world line is to be timelike (or null). In order the world line might be
spacelike, the external field is to be introduced under sign of radical in (%
\ref{a10.1}). Then the expression under sign of radical may be positive even
in the case, when $g_{ik}\dot{q}^{i}\dot{q}^{k}<0$. I introduced the
external field under the sign of radical, writing the action in the form 
\begin{equation}
{\cal A}\left[ q\right] =\int \left\{ -mc\sqrt{g_{ik}\dot{q}^{i}\dot{q}%
^{k}-\alpha f\left( q\right) }+\frac{e}{c}A_{i}\dot{q}^{i}\right\} d\tau ,
\label{a10.2}
\end{equation}
where $f$ is an external scalar field, and $\alpha $ is a small parameter,
which tends to zero at the end of calculations. At the properly chosen field 
$f$ the expression under the radical can be positive even at $g_{ik}\dot{q}%
^{i}\dot{q}^{k}<0$. It appeared that at the properly chosen field $f$, the
world line turned in time indeed. This turn is conserved at $\alpha
\rightarrow 0$. The direct calculations showed that at such a description
the particle energy was positive always, but the time component $p_{0}$ of
the canonical momentum and the particle charge $Q=e{\rm sgn}(\dot{q}^{0})$
depended on sign of derivative $\dot{q}^{0}$, i.e. they were different for
particle and antiparticle. It was rather sudden that the WL charge $Q$,
defined as a source of the electromagnetic field by the relation $Q=\int
\delta {\cal A}/\delta A_{0}({\bf x})d{\bf x}$, did not coincide with the
constant $e$, incoming to the action, although at the correct description
this was to be just so, because the particle and antiparticle had opposite
sign of the charge. One can obtain coincidence of energy $E$ and $p_{0}$, if
one cuts the whole world line into segments, responsible for particles and
antiparticles, and changes direction of the parameter $\tau $ increase on
the segments, responsible for antiparticles, remaining $\tau $ without a
change on segments, responsible for particles. Any change of the sign of $%
\dot{q}^{0}$ is to be accompanied by a change of the sign of the constant $e$%
. But the constant $e$ is a parameter of the dynamic system, and a change of
the sign of $e$ on the segments, describing antiparticles means that
particles and antiparticles are considered to be described in terms of
different dynamic systems.

This simple example shows, that there are two possibilities of description

\noindent (1) To consider the world line (WL) to be a physical object. Then
particle and antiparticle are two different states of WL, distinguishing by
signs of the charge $Q$ and those of the canonical momentum component $p_0$.
The energy is positive in both cases, so $E\ne H$, in general.

\noindent (2) To consider the particle and the antiparticle to be different
physical objects, described by two different dynamic systems. The parameter $%
e$ is the particle charge, which is simultaneously a parameter of the
dynamic system. It is different for a particle and for an antiparticle. At
such a description the evolution parameter $\tau $ can be chosen in such a
way, to provide fulfillment of the constraint $E=H$.

Imposition of the constraint $E=H$ provided automatically fragmentation of
the world line into particles and antiparticles, describing them as
different physical objects, i.e. in terms of different dynamic systems. This
was valid in classical physics. This must be valid in the quantum theory.

It was unclear for me, what was a use of the identification of energy with
Hamiltonian. Why does one cut WL to obtain indefinite nonconservative number
of particles and antiparticles instead of fixed number of physical objects
(WL)? From the formal viewpoint it is more convenient to work with constant
number of object, than with alternating number of them. It was evident for
me, that impossibility of working in QFT without the perturbation theory was
connected directly with the fact that numbers of particles and antiparticles
were not conserved separately. What for does one need to impose the
condition $E=H$ and to restrict one's capacity, if one could impose no
constraints? (Then I did not consider, that the condition $E=H$ might appear
to be incompatible with dynamic equations).

It was necessary to discuss the paper with colleagues dealing with QFT, and
I submitted my report to seminar of the theoretical department of the
Lebedev Physical Institute, where there were many good theorists. At my
report at the session I was surprised by the following circumstance. Nobody
believed that the pair production effect could be described in terms of
classical mechanics. Although my calculations were very simple, they casted
doubt on their validity. It was decided to transfer my report to next
session. One of participants of the seminar (V.Ya.~Fainberg) was asked to
verify my calculations and to report on the next session together with
continuation of my report. Mistakes in my calculations were not found, and I
completed successfully my report on the next session. After the session I
seemed that the attention of participants of the seminar was attracted to
the problem of possibility of pair production description in terms of
classical physics, whereas the main problem, i.e. application the constraint 
$E=H$ in QFT, remained outside the scope. Corresponding my paper was
published \cite{R70}, but, as far as know, nobody payed any attention to it.

It was necessary to quantize nonlinear relativistic field without a use of
the condition $E=H$ and to verify, if such a way of quantization had
advantages over the conventional way, using this condition. It happened that
such a quantization could be carried out without a use of normal ordering
and perturbation theory \cite{R72}. The vacuum state appeared to be
stationary. A possibility of quantization without the perturbation theory
impressed. But I shall not be cunning and say directly, that I had no
illusions about results of my work. In that time (beginning of seventieths)
I assumed that the problem of the quantum mechanics relativization (i.e.
unification of quantum theory with the relativity theory) had no solution. I
assumed that the quantum mechanics was something like relativistic Brownian
motion, and the relativistic quantum theory should be developed in direction
of statistical description of this relativistic motion \cite{R71}.

My work on the secondary quantization of the nonlinear relativistic field
was undertaken with the goal to manifest that the conventional way of the
QFT development was a way to blind alley. The logic of my action was as
follows. One quantizes the nonlinear field, using only principles of
nonrelativistic quantum mechanics and ignoring any additional suppositions.
One advances as far as possible. There were a hope that the quantization
without the perturbation theory admitted one to clarify real problems of QFT
and, maybe, to solve some of them.

The fact was that the use of the perturbation theory did not permit one both
to state exactly problems of QFT and to solve them. The problems of
collisions were the main problems of QFT. To state the collision problem, it
was necessary to formulate exactly what was a particle and what was an
antiparticle. According to quantum mechanics principles it is necessary for
this to define the operator $N_{{\rm p}}^{i}$ of the 4-flux of particles and
the operator $N_{{\rm a}}^{i}$ of the 4-flux of antiparticles. After such a
definition one can state the problem of collisions. Surprisingly, it
appeared that nobody tried to introduce these operators. Instead of this
there were cloudy consideration about the interaction cut off at large time $%
t\rightarrow \pm \infty $. Thereafter these consideration about cut off were
substituted by manipulations with $in$- and $out$-operators, that did not
clarify the statement of the collision problem.

Even in the excellent mathematically rigorous book by F.A.~Berezin \cite{B65}
the collision problem was stated in terms of perturbed $H$ and nonperturbed $%
H_{0}$ Hamiltonians of the system, that corresponds to interaction cut off
at $t\rightarrow \pm \infty $. Of course, all this was only a reflection of
the whole situation in QFT. I asked my colleagues dealing with QFT, how
could one think in terms of the perturbation theory. They answered
obscurely. I understood, that some problems could not be solved exactly. I
was ready to use any methods of approximation (including the perturbation
theory) by the indispensable condition, that the problem be stated exactly,
but not in approximate terms. To state a problem in approximate concepts and
terms was beyond my understanding.

As soon as the nonlinear field was quantized \cite{R72}, results of my paper
were reported on a session of the seminar of the theoretical department of
Lebedev Physical Institute. Although the secondary quantization was produced
without the perturbation theory, most of participants considered my results
to be unsatisfactory on the ground that at the quantization one violated the
condition 
\begin{equation}
\left[ \varphi \left( x\right) ,\varphi ^{\ast }\left( x^{\prime }\right) %
\right] _{-}=0,\qquad \left( x-x^{\prime }\right) ^{2}<0  \label{a10.9}
\end{equation}
which was interpreted usually as the causality condition. Indeed, if at the
quantization the condition $E=H$ is not imposed, the commutator between the
dynamic variables at the points, separated by a spacelike interval $%
x-x^{\prime }$ cannot (and in some cases must not) vanish. Let me explain
this in the example of pair production, described in terms of classical
physics, where the pair production is described by time zigzag of the world
line. In this case the commutator (\ref{a10.9}) associates with the Poisson
bracket. If the condition $E=H$ is imposed and the quantization is carried
out in terms of particles and antiparticles, the dynamic variables $X$ and $%
X^{\prime }$ at the points, separated by a spacelike interval $x-x^{\prime }$%
, relate to different dynamic systems always. The corresponding Poisson
bracket $\{X,X^{\prime }\}$ between any dynamic variables $X$ and $X^{\prime
}$ at these points vanishes. In the case of quantization in terms of world
lines the dynamic variables $X$ and $X^{\prime }$ at the points, separated
by a spacelike interval $x-x^{\prime }$, can belong to the same world line,
i.e. to the same dynamic system. Then the variables $X$ and $X^{\prime }$
correspond to different values $\tau $ and $\tau ^{\prime }$ of evolution
parameter $\tau $. In this case the dynamic variables $X$ at the point $x$
are expressed via dynamic variables $X^{\prime }$ at the point $x^{\prime }$%
, and there exist such a dynamic variables $X_{1}$ at $x$ and $X_{2}^{\prime
}$ at $x^{\prime }$, that the Poison bracket $\{X_{1},X_{2}^{\prime }\}$
does not vanish. The condition (\ref{a10.9}) is violated with a necessity.

Thus, a fulfillment or a violation of the condition (\ref{a10.9}) is an
attribute of a description. It coincides with the causality condition (i.e.
with the objectively existing relation) only at imposition of the condition $%
E=H$. Unfortunately, I failed to convince my opponents of dependence the
relation (\ref{a10.9}) on the way of description, although I tried to do
this at the session and in discussions thereafter. Later on I had
understood, that in this case one met associative delusion, when the
properties of description are attributed to the object in itself.
Unfortunately, it happens that many researchers meet difficulties at
overcoming of AD, and as I am understanding now, the P-style used by the
most researchers of QFT is a reason of these difficulties. Besides,
formulating the condition (\ref{a10.9}) in terms of quantum theory, it is
very difficult to discover that this condition is an attribute of a
description, but not a causality condition.

Thus, I had overcame AD.7, but the scientific community as whole had not
overcame it. I did not see a necessity in further convincing my colleagues
to refuse from imposition of the condition $E=H$ at quantization. At first,
I was convinced that the refusal itself from $E=H$  did not solve main
problems of QFT. My belief, that QFT did not enable to solve the unification
problem of quantum theory with relativity and that the statement of this
problem was false in itself, became stronger. Secondly, I myself did not
know exactly what must replace this problem of unification. I had only a
guess on this account. I might not to convince a person, dealing with QFT
and devoting essential part of his life to this, that he had chosen a wrong
way. Without pointing a right way, such a convincing was useless and even
cruel.

There were once more an important circumstance which influenced strongly on
my interrelations with colleagues dealing with QFT. The fact is that, since
I had discovered incorrectness of imposition the condition $E=H$, I met
difficulties at reading papers on QFT. When I began to read any paper and
discovered that the condition $E=H$ was used there (this was practically in
all papers on QFT), my attention was cut off subconsciously, and I could not
continue conscious reading. My reading became absent-minded, and I needed to
bend my every effort to turn on my attention and continue a conscious
reading. I do not know to what extent such a reaction is my individual
property, but tearing off the papers using $E=H$ led gradually to my allergy
to reading of papers on QFT. I stopped to read them, although I was
interesting QFT always, and questioned my colleagues about QFT development
at any suitable case.

Why did I overcome associative delusions comparatively easy? Apparently, it
was connected with that I was an adherent of the C-style and ignored
instinctively approaches, which were used by the P-style. It is difficult
for me to say, whether this adherence to the C-style was innate, or it was a
result of my education. But such an adherence took place undoubtedly, and
the following case justifies this.

In the beginning of seventieths I had a position of the scientific secretary
of the Space Research Institute. At this position I had a possibility to
investigate all those problems which I wanted, and I dealt with problems of
quantum mechanics and QFT. Thereafter I left this position and passed to the
position of a senior scientific researcher in the department of G.I.~Petrov,
who was a director of the institute. Then I should deal with problems
connected to some extent with space research. I had a possibility of
choosing a field of investigations with one constraint. In the course of a
half a year I was to study literature on the subject, chosen by me and read
a report on the seminar of G.I.~Petrov to manifest my readiness for
investigations in the new field.

Any field of physics, connected with the space research, was new for me, and
I had chosen the problem of the pulsar magnetosphere model and investigation
of the pulsar mechanism. In seventieths it was a slightly developed and
perspective field of astrophysics. The pulsar phenomenon was discovered
recently (in 1967), and there were the first considerations, concerning
pulsar emanation formation. Studying the recent literature on pulsars, I had
read a report on the session of the seminar. Despite the fact that I was a
novice and amateur in this field, my report was very critical. It was
accompanied by a suggestion of my own investigation program.

My pretensions to the authors of reviewed papers were in that they invented
different hypotheses to avoid difficult calculations and attempted to guess
a construction of the pulsar magnetosphere. In my opinion, they should to
state the problem of the pulsar magnetosphere and to solve it by means of
well known methods of classical physics. I considered that hypotheses might
be suggested either on the first (heuristic) stage of investigations (but it
was passed), or after all known methods had been used already. The problem
was in the scope of the classical physics, and conceptual obstacles for its
statement were absent. But most of authors did not want or was not able to
state the problem correctly. Instead of this they advanced their hypotheses,
attempting to explain the pulsar phenomenon at once, and argued, whose
hypothesis is better. In other words, they used the P-style at the
condition, when it was not effective, and it was no necessity of using it.

I did state simply the problem on the pulsar magnetosphere in such a way, as
any researcher, using the C-style in his investigations, should do. I was
working according this program in the course of 1976 -- 1987. (see my
concluding paper on the pulsar model \cite{R88}). In that time I did not
think on styles of investigations. I assumed simply, that one needed to
investigate a physical phenomenon honestly, but not to dodge, substituting
calculations by conjectures. Maybe, my instinctive adherence to the C-style
was so large, that penetrated to my subconsciousness and generated allergy
to reading papers on QFT.

Maybe, me successes in overcoming of different ADs was conditioned by
consecutive application of C-style, essence of which could be expressed by
the Newton's words: ''I do not invent hypotheses''.

\section{Can a curve be a fundamental object \newline
of geometry? Evidence of participant \newline
on completion of the AD.4 overcoming}

My work with technique, based on the world function, showed, that such a
description of geometry was insensitive to topological properties of the
space. For instance, the metric tensor was the same for the Euclidean plane
and for the cylinder, obtained from a band of this plane as a result of
gluing its edges. But the world functions were different. This showed that
the world function describes both local and global properties of geometry.
Further I had discovered, that if all points of the space except for points
with integer value of coordinates removed, the remaining points formed a
discrete geometry, which could be described in terms of the world function.

I understood also, that the world function contained complete information on
geometry. In particular, in the case of the Euclidean space I could obtain
information on the space dimension and construct the Cartesian coordinate
system in the space, even in the case, when the space was an abstract set of
points (but not a manifold). At such a construction one used a concept of
the straight segment between points $P$ and $P^{\prime }$, defined as set of
points $R$, satisfying the equation 
\begin{equation}
S\left( P,R\right) +S\left( R,P^{\prime }\right) =S\left( P,P^{\prime
}\right)  \label{a11.1}
\end{equation}
where $S(P,P^{\prime })$ denotes a distance between the points $P$ and $%
P^{\prime }$. I understood, that the world function of the Euclidean space
possessed special extremal properties. It is these extremal properties that
the equation (\ref{a11.1}) describes a one-dimensional line even in $n$%
-dimensional space, although at an arbitrary function $S(P,P^{\prime })$ the
equation (\ref{a11.1}) describes, in general a $(n-1)$-dimensional surface.
However, then I did not understood, how important was this definition, where
segment of the straight was defined without a reference to the concept of a
curve.

In my representation of that time the world function was a result of
integration of some expression, determined by the metric tensor. This
integration provided a connection between different points of the space.
This connection was conserved, even if the most of space points were removed.

The property of the world function to provide a connection between points
and its extremal properties seemed to me important. Realization of this fact
took place in the middle of sixtieths. I wanted to write a paper about this
properties, but it was not clear, where I could to publish such a paper. The
fact is that this result in itself without applications seemed to me
insufficiently important for its publication.

Opportunity for publication appeared only in the end of eightieths. I
succeeded to publish two my articles, having no relation to geometry \cite
{R89,R989}, in Journal of Mathematical Physics. At the publication of these
articles I was pleasantly surprised by the kind relation of editors. Then I
got an idea of publishing a paper on extremal properties of world function
which reflected my increasing understanding of the world function role in
geometry.

The paper was entitled ''Extremal properties of Synge's world function and
discrete geometry.'' It was accepted to publication, but I was asked, first,
to add references to contemporary papers and, second, to replace designation
''$G$'', which was used for the world function, by conventional designation $%
\sigma $. Addition of reference to more recent papers did not change the
paper practically. The world function was applied in papers on quantization
of gravitation, where ''long-range properties'' of the world function were
important. But nothing except for expansion over degrees of $x-x^{\prime }$
was used in these papers.

But the formal replacement of designation $G$ by another designation $\sigma 
$ led suddenly to essential revision of the paper. The fact is that there
were practically no personal computers in USSR, and my paper was written by
means of typewriter. To change designations I was forced to rewrite the
whole paper. As far as the paper should be retyped, I decided to include all
corrections and additions which appeared in the time, when the paper was
reviewed. In the course of revision I realized the importance of the
definition (\ref{a11.1}) of the straight segment and revised the character
of presentation. Essentially a paper on a construction of a new geometry
appeared, although the term T-geometry was not used in the paper. I began to
think about a change of the title, because the title of the paper did not
reflect its content. The title should be changed.

In that time (1989) it was not simple. To submit a paper in a foreign
journal one needed a permission of ''Glavlit'', where the title of the paper
was mentioned. To obtain a permission for forwarding a manuscript with other
title, it was necessary to translate the new version to Russian and submit
to ''Glavlit''. The procedure of obtaining a new permission would need
one-two months. Besides the paper was to be translated to Russian and typed.
To avoid lack of time, I decided not to change the title. The paper was
printed with the previous title \cite{R90}.

At the transition from the Riemannian geometry to T-geometry it was very
important to overcome inertia of thinking and realize, that a tube could
play a role of the straight. Apparently, physical models of elementary
particles helped me in overcoming of the inertia of thinking. In some models
the particle was substituted by a string or by a membrane, and their world
tubes were respectively two- and three-dimensional surfaces. It meant that
in the real world a tube could play a role of a straight. Psychologically it
was a very important step in construction of T-geometry. Another important
circumstance was the fact that the straight (i.e. a natural geometrical
object, determined by two different points) was defined as a set of points
determined by the metrical property (\ref{a11.1}), i.e. without a reference
to the concept of a curve. In Riemannian geometry the geodesic is defined as
the shortest curve. At such a definition of geodesic, it cannot be a tube,
and a use of the concept of a curve at the geometry construction
discriminates nondegenerate geometries, i.e. geometries, where a tube plays
a role of a straight. Note that the paper \cite{R90} had a transient
character in the sense, that it described rather a transit from the
Riemannian geometry to T-geometry, than the T-geometry as such.

The T-geometry appeared at first as a generalization of the Riemannian
geometry and was applied as a possible space-time geometry \cite{R91}.
Advantages of the space-time model based on T-geometry were evident. First,
it appeared that the geometry depended only on interval between events and
was insensitive to the space-time continuity, which cannot be tested.
Second, in such a geometry the particle mass was geometrized, that was
impossible in the Minkowski geometry. Finally, the stochastic particle
motion, depending on the particle mass (i.e. on geometry), appeared in a
natural way. All this testified that T-geometry was suitable for the
space-time description.

The two-metric technique, developed for the Riemannian space worked very
well in T-geometry, admitting to investigate it in the small \cite{R92}. But
from the practical viewpoint it was unessential, because the behavior of
world lines of particles with a finite mass depended on behavior of the
world function at finite (but not at infinitesimal) intervals.

Although together with construction of T-geometry AD.4 was overcame in
reality, realization of this fact was absent. At first, the T-geometry was
derived as a generalization of Riemannian geometry, and the Riemannian
geometry was considered to be a special case of T-geometry. In particular,
in the paper \cite{R90} one investigated the question, what world function
must be, for degeneration of world tubes into one-dimensional curves would
take place and one could construct a manifold.

At first, I assumed that if the world function was given on a manifold and
satisfied differential equations, which should be satisfied by the world
function of a Riemannian space, then the T-geometry constructed on the base
of this world function be a Riemannian geometry. In reality it was not so,
because the role of the concept of a curve at the construction of the
Riemannian geometry was not exhausted by a construction of geodesics. In the
Riemannian geometry the parallel transport of a vector is founded also on
the concept of a curve.

My interest to geometry was always of an applied character. I interested in
geometry as a method of description of the space-time properties. I did not
thought on a geometry as such, it was for me only a tool, and I troubled
only with efficiency of this tool. In accord with my style of investigations
I did not read papers on geometry, and I troubled only one question whether
I rediscovered any known results. Before publication of the paper \cite{R90}
I had consulted by a well known mathematician, who said that nobody in USSR
dealt with problems close to problems of T-geometry. After publication of
the paper I took several attempts to discuss T-geometry with mathematicians.
The mathematicians, interested in geometry in a large, discussed readily
problems of T-geometry, and I had read several successful reports on
T-geometry. As an ''inventor'' of T-geometry I was very glad, that the
T-geometry was not built before, and I became gradually to understand, what
was a reason of this circumstance. AD.4, i.e. belief of mathematicians that
the curve was a fundamental object of geometry, was a reason of
impossibility of the T-geometry construction.

My reports on seminars of mathematicians were very valuable for me. Firstly,
I had reduced my geometrical ignorance. Secondly, I became gradually to
understand, that T-geometry was rather a generalization of metric geometry,
than that of Riemannian. I discovered that in contemporary mathematics the
geometries did not classified practically, or such a classification was
produced over accidental features. Moreover, there were a lack of
coordination even in the definition, what is the geometry. For instance,
well known mathematician Felix Klein \cite{K37} defined a geometry as a
conception, containing some symmetry group. The Riemannian geometry did not
fall definitely under Klein's definition. According to Klein the Riemannian
geometry should be called the Riemannian topography (or geography).
A.D.~Alexandrov \cite{A48} used another definition of geometry.

The problem of classification is one of the most important problems in any
science, and in mathematics especially. But I did not interested in this
problem, because of my applied approach to geometry. I met this problem at
the following circumstances. On one hand, T-geometry was a very general
geometrical construction, founded only on the world function. Topological
properties appeared to be derivative ones with respect to metric properties
of geometry. From viewpoint of T-geometry one cannot set topological
properties of the space independently. On the other hand, in the generalized
Riemannian geometry \cite{T59, ABN86,BGP92} the topological properties are
introduced at first, and thereafter one introduces metrical properties, what
one cannot make from the viewpoint of T-geometry. In general, as far as I
could understand, in present time the topology is considered to be the most
promising direction of the geometry development. In support of this
statement one can mention the following fact. In the faculty of mechanics
and mathematics of Moscow Lomonosov university there are three different
chairs, whose titles contain words ''geometry'' and ''topology'' in
different combinations.

As far as I saw a contradiction between the existence of T-geometry and the
topology as the most promising direction in geometry, the situation should
be clarified in the process of discussions. My attempts to report T-geometry
on seminars of the ''most geometrical'' chairs failed. Corresponding
suggestions were declined on the ground that such a report would not be
interesting for researchers of these chairs, who were interested in other
problems. Then I prepared a report, entitled ''Geometry without topology''
and submitted it to Moscow Mathematical Society (MMS) with a request to hear
it on one of sessions.

MMS (a very respectable organization) heard on its sessions only very
important reports (preliminary reviewed by experts). Review of my paper was
negative. My work on T-geometry was recognized by the reviewer as
insufficiently fundamental and unworthy of hearing on a session of MMS. This
was not quite that, what I needed, because I needed a public discussion.
Nevertheless, my goal was attained, although only by half. I obtained
opinion of the expert, who was mostly competent in this problem. I had
clarified from his review, that there are no classifications of geometries
in mathematics, and the problem of ''fundamental nature'' of geometry in
question is determined entirely by opinion of a reviewer.

My argued appeal was not taken into account, but the other appeared to be
important. It was important, that the negative review and reviewer's
argumentation forced me to think about the problem of ''fundamental nature''
of a geometry. As a result I came to a necessity of classification of
geometries, and a key to the classification was the concept of CG
(conception of geometry) as a method of a construction of a standard
(Euclidean) geometry. A criterion of generality of CG was the amount of
numerical information, which the given CG used for construction of the
standard geometry. Such a classification of CG is presented on p.~\pageref
{tab}. It follows from this classification that purely metric CG generates a
class of the most general geometries (T-geometries), whereas perspectives of
topology-metric conception are unfavourable in this respect. The review was
dated by the seventh November 2000. A result of my work over this review was
a revision of my report, which turned to the article ''Geometry without
topology as a new conception of geometry'' \cite{Rb001}. The paper was ready
to the end of 2000th year. This paper was a corollary of my complete
overcoming of AD.4.

Now I am presenting chronology of my overcomings of AD.4 -- AD.7 and trying
to find the reasons, why I have succeeded to overcame these associative
delusions. AD.5 -- AD.7 were overcame practically in the same time in the
beginning of seventieths of XX century. This overcoming was a corollary of
my understanding of the simple statement, that in the relativistic physics
the physical object is WL (world line), but not a particle. Overcoming of
AD.5 and AD.7 followed directly from this circumstance. Overcoming of AD.6
was a simple corollary of the fact, that the quantum mechanics is a
statistical description of stochastic relativistic particles motion. Thus,
finally, overcoming of AD.5 -- AD.7 was {\it a corollary of my understanding
of the relativity theory}. Overcoming of AD.4 was happening in the course of
a long time. It was finished in the end of eightieths of XX century. Then it
appeared in the form of a guess. This overcoming in the form of a logical
corollary happened in the end of 2000 year, when {\it the conception of
geometry} was introduced and the table, presented on p.~\pageref{tab} was
obtained..

What is a reason of these overcoming? What properties did I have and did not
have other researchers? I believe that these properties were as follows:

\noindent (1) My understanding of the relativity theory. I thought in terms
of space-time diagrams, i.e. geometrically, but not in terms of relativistic
invariance as most of researchers. I am connecting such a way of thinking
with that, I had studied the relativity theory on the book of V.A.~Fock \cite
{F55}.

\noindent (2) In my investigations I used the C-style and was consecutive in
my research. The logical consistence of a theory was more important for me,
than its practical results.

\noindent (3) In all time of my investigation activity I had not in reality
any scientific divisor, i.e. there were no sufficiently authoritative for me
person, who would say me, that I should use the more pragmatic P-style
instead of C-style, that I used.

\noindent (4) I was sufficiently self-reliant, to follow the way, that I was
led by the logic. I did not pay attention to other researchers, ignored all
authorities and the circumstance, so that some researchers considered me as
a scientific dissident.

I present the peculiarities, which seems for me to be necessary for
overcoming of associative delusions. Each of this peculiarities appears
rather rare. Collection of all these peculiarities at one person is the more
rarity.

To describe my research activity briefly, one should say, that {\it using
C-style,} I {\it put consecutively into effect the idea of geometrization of
physics,} and this agreed completely with the general line of the physics
development in XIX -- XX centuries.

\section{Concluding remarks}

Thus, the associative delusions (AD) accompanied the cognition process.
Although one should tend to eliminate ADs, but, apparently, the complete
elimination of them is impossible. In the case of impossibility of this
elimination of ADs, AD leads to appearance of additional compensating
hypotheses and to a construction of compensating (Ptolemaic) conceptions.
Appearance of Ptolemaic conceptions leads to a generation of a special
P-style of investigations, suitable for work with Ptolemaic conceptions. The
P-style is simultaneously a style of investigations and a style of thinking.
On one hand, the P-style is ''noise-resistant'' (suitable for work with
Ptolemaic constructions, containing false suppositions), but on the other
hand, it is less predictable, than C-style. In the course of some time one
can pursue investigations, using P-style. But, thereafter the Ptolemaic
conceptions stops to be effective. It becomes necessary to find and to
overcome corresponding AD, returning to C-style. If the P-style was existing
for a long time and several generations of researchers had educated on its
application, the overcoming of AD and returning to the C-style will be a
difficult process. One needs to be ready to this.

After discovering AD the subsequent revision of existing theory may appear
to be very essential. If it concerns the space-time geometry, the revision
may lead even to a change of a world outlook. Transition from the space-time
with the primordially deterministic particle motion to the space-time with
the primordially stochastic motion is already a ground for a change of the
world outlook. If earlier it was necessary to explain the stochasticity,
starting from the determinism of the world, then now one should explain
deterministic phenomena on the basis of primordial stochasticity of the
world.

\end{document}